\documentstyle[12pt,epsfig]{article}
\newcommand{\be}{\begin{equation}}
\newcommand{\ee}{\end{equation}}
\newcommand{\beq}{\begin{equation}}
\newcommand{\eeq}{\end{equation}}
\newcommand{\ba}{\begin{eqnarray}}
\newcommand{\ea}{\end{eqnarray}}
\newcommand{\bea}{\begin{eqnarray}}
\newcommand{\eea}{\end{eqnarray}}

\begin{document}
\baselineskip=15.5pt \pagestyle{plain} \setcounter{page}{1}

\begin{titlepage}

\vskip 0.8cm

\begin{center}

{\LARGE Deep Inelastic Scattering from Holographic Spin-One Hadrons}
\vskip .3cm

\vskip 1.cm

{\large {Ezequiel Koile, Sebastian Macaluso and Martin
Schvellinger} }

\vskip 1.cm

{\it IFLP-CCT-La Plata, CONICET and \\
Departamento  de F\'{\i}sica, Universidad Nacional de La Plata.
\\ Calle 49 y 115, C.C. 67, (1900) La Plata,  \\ Buenos Aires,
Argentina.} \\

\vspace{1.cm}

{\bf Abstract}

\end{center}

We study deep inelastic scattering structure functions from hadrons
using different holographic dual models which describe the strongly
coupled regime of gauge theories in the large $N$ limit.
Particularly, we consider scalar and vector mesons obtained from
holographic descriptions with fundamental degrees of freedom,
corresponding to ${\cal {N}}=2$ supersymmetric and
non-supersymmetric Yang-Mills theories. We explicitly obtain
analytic expressions for the full set of eight structure functions,
{\it i.e.}, $F_1$, $F_2$, $g_1$, $g_2$, $b_1$, $b_2$, $b_3$, $b_4$,
arising from the standard decomposition of the hadronic tensor of
spin-one hadrons. We obtain the relations $2 F_1 = F_2$ and $2 b_1 =
b_2$. In addition, we find $b_1 \sim {\cal {O}}(F_1)$ as suggested
by Hoodbhoy, Jaffe and Manohar for vector mesons. Also, we find new
relations among some of these structure functions.

\noindent

\end{titlepage}

\newpage

\section{Introduction}

There are holographic dual models based on type IIA and type IIB
string theories which describe certain aspects of strong
interactions when matter fields transforming under the fundamental
representation of the gauge group are taken into account
\cite{Kruczenski:2003be,Sakai:2003wu,Babington:2003vm,Kruczenski:2003uq,Sakai:2004cn}.
In these D-brane constructions mesons are modeled as fluctuations of
flavour branes in the probe approximation. From some of these models
it is possible to derive properties of mesons such as ratios of
masses, decay constants and the chiral Lagrangian, among others.
This makes the alluded models interesting for investigating certain
aspects of the phenomenology of strong interactions. Mesons derived
from this kind of models are generically known as holographic
mesons. One further important and very interesting question that one
may ask in the context of these models is about the structure of the
holographic mesons, when they are probed by virtual photons in deep
inelastic scattering (DIS) processes.

The DIS cross section is proportional to the Lorentz contraction of
a leptonic tensor, entirely computed within perturbative quantum
field theory, and a hadronic tensor $W_{\mu\nu}$ which receives
contributions from the strong coupling regime of QCD. The hadronic
tensor is computed using the optical theorem, and is recast in terms
of the two-point correlation function of electromagnetic currents
inside the scattered hadron. From the most general Lorentz
decomposition of this tensor the so-called structure functions can
be extracted. In this work we shall be focussed on DIS from mesons.
Spin-zero mesons have only two structure functions called $F_1$ and
$F_2$. In addition, polarized spin-one hadrons have eight structure
functions \cite{Hoodbhoy:1988am}, as we shall review in the next
section.

From the point of view of the gauge/string duality
\cite{Maldacena:1997re,Gubser:1998bc,Witten:1998qj}, Polchinski and
Strassler \cite{Polchinski:2002jw} proposed a method for deriving
the hadronic tensor and the DIS structure functions for glueballs
and spin-$\frac{1}{2}$ hadrons, focussing on confining gauge
theories that are conformal (or nearly conformal) at momenta well
above the confinement scale $\Lambda$. For instance, one can think
of the ${\cal {N}}=1^*$ SYM theory
\cite{hep-th/9510101,hep-th/0003136}, that can be obtained from the
${\cal {N}}=4$ SYM theory explicitly broken at a certain mass scale
to pure ${\cal {N}}=1$ SYM theory. Further developments have applied
the ideas of \cite{Polchinski:2002jw} to scalar mesons, unpolarized
vector mesons \cite{Bayona:2009qe,BallonBayona:2010ae} and nucleons
\cite{Gao:2009ze,Gao:2010qk}. On the other hand, DIS structure
functions \cite{Hatta:2007cs} and lepton-pair photo-production rates
\cite{CaronHuot:2006te} have also been studied in the strong
coupling regime of an ${\cal {N}}=4$ SYM plasma using a holographic
dual description to obtain correlation functions of two
electromagnetic currents. More recently, it has been considered
${\cal {O}}(\alpha'^3)$ corrections from type IIB string theory on
DIS \cite{Hassanain:2009xw}, plasma photoemission
\cite{Hassanain:2011ce}, as well as other plasma properties which
rely on the computation of two-point correlation functions of
electromagnetic currents such as the electrical conductivity
\cite{Hassanain:2011fn, Hassanain:2010fv}.

In this paper we investigate DIS from polarized spin-one mesons and
obtain the DIS structure functions by extending the proposal of
\cite{Polchinski:2002jw}. We derive the interactions in the bulk
directly from the expansion up to quadratic order in the derivatives
of the fields of the Dirac-Born-Infeld action of the flavour branes
in the probe approximation for the $\textmd{D3-D7}$-brane model
\cite{Kruczenski:2003be} and the
$\textmd{D4-D8-}\overline{\textmd{D8}}$-brane model
\cite{Sakai:2004cn}. The derivation we use, which is very different
from the one presented in \cite{Bayona:2009qe,BallonBayona:2010ae},
is totally general and can be applied to other holographic dual
models and for different holographic hadrons. We also obtain the
corresponding DIS structure functions for scalar mesons using these
models.

The main difference with \cite{Bayona:2009qe,BallonBayona:2010ae} is
the way we construct the interaction in the bulk. So, let us
summarize the method we use in this paper. The holographic dual
models we consider have an internal Einstein manifold which contains
a sphere\footnote{In a more general holographic dual model the
sphere could be replaced by any other compact Einstein manifold, and
the same argument holds.}: {\it i.e.}  the D3-D7 brane model has an
$S^3$ and the Sakai-Sugimoto model has an $S^4$. The isometry groups
are $SO(4)$ and $SO(5)$, respectively. We can consider a $U(1)$
subgroup of these isometry groups. From the point of view of the
dual gauge theory, these global bosonic symmetries correspond to the
$R$-symmetry group $SU(2)_R$ in the D3-D7 brane model, while for the
Sakai-Sugimoto model this corresponds to the global symmetry group
$SO(5)$. Now, consider a field in the probe-brane world-volume in
each of these models. Its wave function can be factorized as the
product of a plane-wave in the four-dimensional Minkowski spacetime
times a function depending on the radial coordinate, times a
spherical harmonic on the corresponding sphere. Notice that this
spherical harmonic satisfies an eigenvalue equation, with eigenvalue
equal to the charge under the referred global symmetry, let us call
it ${\cal {Q}}$. Also, recall that for each global continuous
symmetry there is a corresponding conserved Noether's current. This
is the situation in the bulk. Now, suppose we insert a $U(1)$
current at the boundary theory, this induces a fluctuation on the
boundary conditions of the bulk fields coupled to this current at
the boundary. The fluctuation propagates in the bulk. The precise
form of this fluctuation comes from the fact that this boundary
current is a Lorentz vector which couples to the boundary value of a
gauge field propagating in the bulk. As we shall explain later, this
is an off-diagonal fluctuation on the bulk metric of the form
$\delta g_{mj}=A_m(y, r) v_j(\Omega)$, where $A_m$ is the gauge
field and $v_j$ is the Killing vector corresponding to the isometry
subgroup $U(1) \subset SO(n+1)$ of the $S^n$ of each considered
model. In this way, it is natural to straightforwardly construct the
interaction Lagrangian in the bulk by coupling the bulk Noether's
current above to the gauge field $A_m$ with a strength given by the
charge ${\cal {Q}}$. So, this is the main point in the construction
that follows here. The existence of the bulk conserved Noether's
current is crucial in this construction.

The main new results we present in this paper can be summarized as
follows. We develop a consistent method to derive the interaction
Lagrangian from the Dirac-Born-Infeld action of flavour probe
branes. Then, we obtain the full set of structure functions for
scalar and polarized vector mesons in both holographic dual models.
In addition, we find a completely new full tower of vector mesons
from the Sakai-Sugimoto model. The reason to use these two very
different holographic dual models with flavours to study deep
inelastic scattering from polarized spin-one hadrons is to be able
to compare the structure functions and discuss their model-dependent
as well as model-independent properties. We have obtained analytic
expressions for the full set of eight structure functions: $F_1$,
$F_2$, $g_1$, $g_2$, $b_1$, $b_2$, $b_3$, $b_4$, which come from the
standard decomposition of the hadronic tensor of spin-one hadrons.
They are functions of two independent dimensionless kinematical
variables: $x$ (the Bjorken parameter) and $t$, both to be defined
in the next section. Indeed, we show that these functions split into
a model-dependent factor, which is common to the eight structure
functions of each model, and a model-independent one. We discuss
model-independent properties of the structure functions, such as the
relations $2 F_1(x) = F_2(x)$ and $2 b_1(x) = b_2(x)$, which we have
obtained for all values of the Bjorken parameter (neglecting terms
proportional to $t$), though the supergravity approximation only
holds for values of $x$ close to one\footnote{Initial gauge/string
duality studies for small values of the Bjorken parameter can be
seen for example at references
\cite{Polchinski:2002jw,Brower:2006ea,Cornalba:2008sp,Cornalba:2009ax,Cornalba:2010vk}.
}. We have found a missing $x$-factor in comparison with the usual
Callan-Gross relations. This is due to the fact that, within the
pure supergravity description, lepton scattering is produced from
the entire hadron \cite{Polchinski:2002jw}. Also, we have found the
relation $b_1(x) = 3 F_1(x)$ in agreement with the expectations from
\cite{Hoodbhoy:1988am} for the $\rho$-meson, as well as an
additional set of new relations among some of these structure
functions that we explain in the last section of the paper.

In Section 2 we describe the kinematics of the DIS from polarized
spin-one mesons which shall be relevant for our studies in the
following sections. In Section 3 we very briefly review the proposal
of reference \cite{Polchinski:2002jw}. In Section 4 we begin with a
brief review of the D3-D7-brane model description following
\cite{Kruczenski:2003be} and we show the expressions for the scalar
and vector mesons. Then, we develop a consistent method to obtain
the interaction Lagrangian between the mesons and the gauge field
which arises from the fluctuation of the metric induced by the
insertion of a current at the boundary of the spacetime. We obtain
the hadronic tensor and the corresponding eight DIS structure
functions from polarized spin-one vector mesons of the dual ${\cal
{N}}=2$ gauge theory. In Section 5, we carry out a similar programme
for the Sakai-Sugimoto model \cite{Sakai:2004cn} using our method to
derive the interaction Lagrangian from first principles. We have
also obtained a full tower of vector fluctuations of the D8-brane,
which has an additional quantum number, $\ell$, compared to the
mesons considered in \cite{Sakai:2004cn}, that comes from the
expansion on spherical harmonics on $S^4$. Besides, due to the fact
that the DIS relevant bulk interaction region corresponds to large
values of the four-dimensional energy, we consider the metric for
that specific region. This allows us to calculate full analytic
expressions for these mesons depending on the nine coordinates on
the D8-brane. We extensively discuss our results in the last section
of the paper.

\section{Kinematics of DIS from spin-one hadrons}\label{Kinematics}

Deep inelastic scattering is a high energy process which allows to
investigate the hadronic structure. Typically, a lepton is scattered
from a hadron in a kinematical regime where the hadron becomes
fragmented in many particles which are not measured. The process can
be described as the electromagnetic scattering of a lepton by a
quark or a parton inside the hadron. In this section we follow the
conventions of Manohar \cite{Manohar:1992tz}, except for the
four-dimensional Minkowski metric that we take as
$\eta_{\mu\nu}\equiv \textmd{diag}(-1,+1,+1,+1)$. So, let us briefly
summarize the kinematics of DIS from vector and scalar mesons
following the quantum field theory analysis from references
\cite{Manohar:1992tz} and \cite{Hoodbhoy:1988am}.

Let us consider an incoming lepton beam with four-momentum $k^\mu$
(where $k^0\equiv E$) which will be scattered from a fixed hadronic
target. The four-momentum of the scattered lepton $k'^\mu$ (with
$k'^0\equiv E'$) is measured, but the final hadronic state (that we
call $X$) is not. The lepton exchanges a virtual photon of
four-momentum $q^{\mu}$ with the initial hadronic state. The virtual
photon probes the hadron at distances as small as $1/\sqrt{q^{2}}$.
If the hadron is not fragmented the scattering is called elastic,
otherwise the resulting process is DIS.

The DIS amplitude is given by:
\begin{equation}
i{\cal{M}}=(-ie)^{2}
\left(\frac{-i\eta_{\mu\nu}}{q^{2}}\right)\langle
k'|J^{\mu}_{l}(0)|k,s_{l}\rangle\langle
X|J^{\nu}_{h}(0)|P,h\rangle\, ,
\end{equation}
where $e$ is the lepton electric charge, $s_{l}$ is the polarization
of the incoming lepton, $h$ is the polarization of the hadronic
initial state, and $J^{\mu}_{l}, \, J^{\nu}_{h}$ are the
electromagnetic currents of the lepton and hadron, respectively. The
polarization $h$ can be chosen as the spin in the direction of an
arbitrary axis, usually the direction of the incident beam. The
differential DIS cross section is then
\begin{eqnarray} \label{DIS3}
d\sigma &=& \sum_{X}\int\frac{d^{3}k'}{(2\pi)^{3}2E'}(2\pi)^{4}
\delta^{4}(k+P-k'-P_{X})\frac{|{\cal {M}}|^{2}}{(2E)(2M)(v_{rel}=1)} \nonumber\\
&=& \sum_{X}\int\frac{d^{3}k'}{(2\pi)^{3}2E'}\frac{(2\pi)^{4}
\delta^{4}(k+P-k'-P_{X})}{(2E)(2M)}\frac{e^{4}}{q^{4}}\\
& & \times \langle P,h|J^{\mu}_{h}(0)|X\rangle\langle
X|J^{\nu}_{h}(0)|P,h\rangle \langle k,s_{l}|J_{l
\mu}(0)|k'\rangle\langle k'|J_{l \nu}(0)|k,s_{l}\rangle \, ,
\nonumber
\end{eqnarray}
where $P^\mu$ and $P_X^\mu$ are the hadronic initial and final
momentum. $M^2=-P^2$ and $M_X^2=-P_X^2$ are the initial and final
hadronic squared masses, respectively. The relation
$\langle\alpha|J^{\mu}|\beta\rangle^{*}=\langle\beta|J^{\mu}|\alpha\rangle$
has been used, since $J^{\mu\dagger}=J^{\mu}$. Notice that since the
polarizations of the final lepton and hadron states are not measured
there is a sum over them.

The leptonic tensor is defined as
\begin{equation}\label{DIS3.1}
l^{\mu\nu}=\sum_{final \ spin}\langle
k'|J_l^{\nu}(0)|k,s_l\rangle \langle
k,s_l|J_l^{\mu}(0)|k'\rangle\,,
\end{equation}
and can be computed pertubatively within quantum electrodynamics.
For a spin-$\frac{1}{2}$ lepton it is given by
\begin{equation}\label{DIS3.2}
l^{\mu\nu}=2[k^\mu k'^\nu + k^\nu k'^\mu -\eta^{\mu\nu}(k\cdot k'-m_l^2)
-i\epsilon^{\mu\nu\alpha\beta}q_\alpha s_{l\beta}],
\end{equation}
where $m_{l}$ is the lepton mass.

On the other hand, the hadronic tensor $W_{\mu\nu}$ is given by \cite{Manohar:1992tz}
\begin{equation}\label{DIS4}
W_{\mu\nu}(P,q)_{h'\,h}=\frac{1}{4\pi}\int d^{4}x\, e^{iq.x}\langle
P,h'|[J_{\mu}(x),J_{\nu}(0)]|P,h\rangle\,,
\end{equation}
where $h$ and $h'$ are the polarizations of the initial and final
hadronic states, respectively. By inserting a complete set of
eigenstates and taking into account translational invariance, the
above expression becomes
\begin{eqnarray}\label{DIS5}
W_{\mu\nu}(P,q)_{h'\,h}=\frac{1}{4\pi}\sum_{X}\bigg[(2\pi)^{4}\delta^{4}(q+P-P_{X})
\langle P,h'|J_{\mu}(0)|X\rangle\langle X|J_{\nu}(0)|P,h\rangle\nonumber \\
-(2\pi)^{4}\delta^{4}(q-P+P_{X})\langle
P,h'|J_{\nu}(0)|X\rangle\langle X|J_{\mu}(0)|P,h\rangle\bigg]\,.
\end{eqnarray}
The allowed final states must satisfy the condition $P_{X}^{0}\geq
P^{0}$, since $M_{X}^{2}\geq M^{2}$. Since $q^{0}>0$, only the first
delta function in Eq.(\ref{DIS5}) fulfills this condition, thus the
sum in $W_{\mu\nu}$ reduces to Eq.(\ref{DIS3}). Finally, one gets
\begin{equation}\label{DIS7}
\frac{d^{2}\sigma}{dE'd\Omega}=\frac{e^{4}}{16\pi^{2}q^{4}}
\frac{E'}{ME}l^{\mu\nu}W_{\mu\nu}(P,q)_{h'\,h}\,.
\end{equation}
It is important to notice from Eq.(\ref{DIS3.2}) that the
spin-independent part of the leptonic tensor is symmetric under
$\mu\leftrightarrow\nu$, while its spin-dependent part is
antisymmetric. Therefore, an unpolarized lepton beam only probes the
symmetric part of the hadronic tensor $W_{\mu\nu}$.

The relevant hadronic structure for DIS can be completely
characterized by $W_{\mu\nu}$. The probability that a hadron
contains a given constituent with a given fraction $x$ of its total
momentum is given by the partonic distribution functions, which
unfortunately cannot be computed with perturbative QCD. This is
because they depend on soft (non-perturbative) QCD dynamics which
determines the hadronic structure as a confined state of quarks and
gluons. Therefore, the incoming lepton gets scattered from the
hadronic constituents which carry a fraction $x$ of the total
hadronic momentum.

For the hadrons composed by massless partons the probability of
finding a parton with a momentum $x P^\mu$ is given by the
distribution function $f(x, q^{2})$. When the partons are free, this
function becomes independent of $q^{2}$, {\it i.e.} $f\equiv f(x)$,
which is the so-called Bjorken scaling. Obviously, this scaling does
not hold for QCD where the partonic distribution functions change as
a function of $q^{2}$, since each parton tends to fragment into
multiple partons with lower $x$. Thus, the hadronic structure in QCD
does depend on $q^{2}$, being the parton number increased while the
average value of $x$ decreases as $q^{2}$ increases.

Now, let us focus upon the hadronic structure functions. First, let
us review a few basic properties. They are dimensionless functions
depending on $P^{2}$, $P \cdot q$ and $q^{2}$. It is useful to write
them in terms of the dimensionless variables $t \equiv
\frac{P^2}{q^2}$ and $x\equiv -\frac{q^2}{2 P \cdot q}$. The ranges
of these variables are $0 < x \le 1$ and $t < 0$. The structure
functions are obtained from the most general Lorentz decomposition
of the hadronic tensor, satisfying certain physical requirements for
$W_{\mu\nu}$. For parity preserving interactions it is necessary to
satisfy current conservation $\partial_{\mu}J^{\mu}(x)=0$ which
implies that
\begin{equation}\label{DIS11}
q^{\mu} W_{\mu\nu}(P, q, h) = q^{\nu} W_{\mu\nu}(P, q, h) = 0 \, .
\end{equation}
It must also be parity invariant:
\begin{equation}\label{DIS13}
W_{\mu\nu}(P, q, h) =
W_{\mu_{\textmd{p}}\nu_{\textmd{p}}}(P_{\textmd{p}}, q_{\textmd{p}},
h_{\textmd{p}})\,,
\end{equation}
where the subindex $\textmd{p}$ indicates parity-transformed
quantities. Time-reversal symmetry must also be satisfied, which
means that
\begin{equation}\label{DIS114}
W_{\mu\nu}(P , q, h) = W^{*}_{\nu_{T}\mu_{T}}(P_{T}, q_{T}, h_{T})\,
,
\end{equation}
where the subindex $T$ means time reversal. In addition, there is an
identity relating $W_{\mu\nu}$ for $q^{0}>0$ with $W_{\mu\nu}$ for
$q^{0}<0$:
\begin{equation}\label{DIS15b}
W_{\mu\nu}(P, q, h)=-W_{\nu\mu}(P, -q, h)\,,
\end{equation}
which comes from translation invariance. At this point we introduce
the hadronic tensor for polarized spin-one hadrons for
parity-preserving interactions as derived in \cite{Hoodbhoy:1988am}
by Hoodbhoy, Jaffe and Manohar. There are four new structure
functions in comparison with spin-$\frac{1}{2}$ hadrons, which are
called $b_{1}$, $b_{2}$, $b_{3}$ and $b_{4}$. They contribute to the
symmetric part of $W_{\mu\nu}$, thus contributing to DIS from an
unpolarized target. Therefore, the hadronic tensor can be written in
terms of eight independent structure functions. Omitting terms
containing $q_{\mu}$ and $q_{\nu}$, since they vanish after the
contraction with the leptonic tensor, the hadronic tensor reads
\begin{eqnarray}\label{DIS16}
W_{\mu\nu} &=& F_{1}\eta_{\mu\nu}-\frac{F_{2}}{P \cdot
q}P_{\mu}P_{\nu}+b_{1}r_{\mu\nu}-\frac{b_{2}}{6}(s_{\mu\nu}+t_{\mu\nu}+
u_{\mu\nu}) -\frac{b_{3}}{2}(s_{\mu\nu}-u_{\mu\nu})\nonumber\\ & &
-\frac{b_{4}}{2}(s_{\mu\nu}-t_{\mu\nu}) -\frac{ig_{1}}{P \cdot
q}\epsilon_{\mu\nu\lambda\sigma}q^{\lambda}s^{\sigma}
-\frac{ig_{2}}{(P \cdot q)^{2}}\epsilon_{\mu\nu\lambda\sigma}q^{\lambda}
(P \cdot q \:s^{\sigma}-s \cdot q \:P^{\sigma})\,,\nonumber\\
\end{eqnarray}
where we have used the following definitions:
\begin{eqnarray}
\label{DIS17} && r_{\mu\nu}\equiv\frac{1}{(P \cdot q)^{2}}\bigg(q
\cdot \zeta^{*}\:q \cdot \zeta-\frac{1}{3}(P \cdot q)^{2}\kappa
\bigg)\eta_{\mu\nu} \, , \\
\label{DIS18} && s_{\mu\nu}\equiv\frac{2}{(P \cdot q)^{3}}\bigg(q
\cdot \zeta^{*}\:q \cdot \zeta-\frac{1}{3}(P \cdot q)^{2}\kappa
\bigg)P_{\mu}P_{\nu} \, , \\
\label{DIS19} && t_{\mu\nu}\equiv\frac{1}{2(P \cdot q)^{2}}\bigg(q
\cdot \zeta^{*}\:P_{\mu}\zeta_{\nu}+q \cdot
\zeta^{*}\:P_{\nu}\zeta_{\mu} +q \cdot
\zeta\:P_{\mu}\zeta^{*}_{\nu}+q \cdot \zeta
\:P_{\nu}\zeta^{*}_{\mu}-\frac{4}{3}(P \cdot q) P_{\mu}P_{\nu}\bigg) \, , \nonumber \\
&& \\
\label{DIS20} && u_{\mu\nu}\equiv\frac{1}{P \cdot
q}\bigg(\zeta^{*}_{\mu}\zeta_{\nu}+\zeta^{*}_{\nu}\zeta_{\mu}
+\frac{2}{3}M^{2}\eta_{\mu\nu}
-\frac{2}{3}P_{\mu}P_{\nu}\bigg) \, , \\
\label{DIS21} &&
s^{\sigma}\equiv\frac{-i}{M^{2}}\epsilon^{\sigma\alpha\beta\rho}
\zeta^{*}_{\alpha}\zeta_{\beta}P_{\rho}
\, ,
\end{eqnarray}
being $\kappa=1 - 4 x^{2} t$ and $s^{\sigma}$ a four-vector
analogous to the spin four-vector for spin-$\frac{1}{2}$ particles.
Besides, $\zeta$ and $\zeta^*$ denote the initial and final hadronic
polarization vectors, respectively. The condition $P \cdot \zeta=0$
is satisfied, and the normalization is given by $\zeta^{2}=-M^{2}$.

Recall that DIS amplitudes can be obtained from the imaginary part
of the forward Compton scattering amplitudes. In particular, from
the matrix element of two electromagnetic currents inside the
hadron, the tensor $T_{\mu\nu}$ is defined as
\begin{equation}\label{DIS50}
T_{\mu\nu}=i\langle P, {\cal {Q}}|{\hat{T}}(\tilde J_{\mu}(q) \,
J_{\nu}(0))|P,  {\cal {Q}}\rangle\,,
\end{equation}
where, as before, $P$ is the four-momentum of the initial hadronic
state (where for brevity we have omitted its Lorentz index), $q$ is
the four-momentum  of the virtual photon and ${\cal {Q}}$ is the
charge of the hadron.
${\hat{T}}(\hat{{\cal{O}}}_{1}\hat{{\cal{O}}}_{2})$ indicates
time-ordered product between $\hat{{\cal{O}}}_{1}$ and
$\hat{{\cal{O}}}_{2}$ operators. The tilde stands for Fourier
transform. The tensor $T_{\mu\nu}\equiv T_{\mu\nu}(P, q, h)$ has
identical symmetry properties as $W_{\mu\nu}(P, q, h)$, thus having
similar Lorentz-tensor structure to $W_{\mu\nu}$. Using the optical
theorem one gets
\begin{equation}\label{DIS51}
\textmd{Im}\,\tilde{F_{j}}=2\pi\,F_{j}\,,
\end{equation}
being $\tilde{F_{j}}$ the $j$-th structure function of $T_{\mu\nu}$
and $F_{j}$ the corresponding one of $W_{\mu\nu}$.

\section{A holographic dual description of DIS}

Let us very briefly describe the idea presented in
\cite{Polchinski:2002jw} by Polchinski and Strassler to study DIS
for gauge theories that have holographic dual descriptions. Full
details are explained in the original reference. Within the
supergravity approximation they calculate hadronic structure
functions for $x\sim1$ (with $x<1$)\footnote{In
\cite{Polchinski:2002jw} also a small-$x$ calculation was carried
out in terms of a string theory analysis, {\it i.e.} beyond the pure
supergravity approximation.}. In particular, they consider confining
gauge theories in four dimensions such as certain deformations of
${\cal{N}}=4$ SYM, from which they study DIS from glueballs and
spin-$\frac{1}{2}$ hadrons. The examples considered in that paper
are UV conformal or nearly conformal. Thus, the dual string theory
is defined on the $AdS_{5}\times W$ background, where $W$ can be an
Einstein manifold. The metric is given by
\begin{equation}\label{pol1}
ds^{2}=\frac{r^{2}}{R^{2}}\eta_{\mu\nu}dy^{\mu}dy^{\nu}
+\frac{R^{2}}{r^{2}}dr^{2}+R^{2}\widehat{ds}^{2}_{W}\,,
\end{equation}
where $R=(4\pi g_s N)^{1/4}\alpha'^{1/2}$ is the $AdS_{5}$ radius
when $W$ is the five-sphere\footnote{Otherwise it must be other
volume factors multiplying the relevant factor $(g_s
N)^{1/4}\alpha'^{1/2}$ of the metric scale, however, for those cases
the present discussion also applies.}. The four coordinates
$y^{\mu}$ are identified with those of the gauge theory, while $r$
is the holographic radial one. Coordinates on $W$ are denoted by
$\Omega$. The ten-dimensional energy scale is given by $R^{-1}$ (up
to powers of the 't Hooft coupling $\lambda = g_{YM}^2 N \equiv 4
\pi g_s N$, with the string coupling $g_s$), while the
characteristic four-dimensional energy is
\begin{equation}\label{pol2}
E^{(4)}\sim\frac{r}{R^{2}}\,.
\end{equation}
If one is interested in the large $N$ limit of confining gauge
theories, the geometry of the holographic dual model at large $r$ is
approximately that of Eq.(\ref{pol1}), however, it must be modified
at a radius corresponding to
\begin{equation}
r \sim r_{0}=\Lambda R^{2}\,,
\end{equation}
where $\Lambda$ is the confinement scale. Nevertheless, the dynamics
of interest for $q\gg\Lambda$ corresponds to $r_{int}\sim qR^{2}\gg
r_{0}$, where the conformal metric (\ref{pol1}) can be used.
$r_{int}$ denotes the bulk region where the relevant interaction
occurs as we explain below.

Polchinski and Strassler employ the dual string theory description
to obtain the matrix element $T^{\mu\nu}$. Its imaginary part is
given by
\begin{eqnarray}\label{pol3}
\textmd{Im} \:T^{\mu\nu} &=& \pi\sum_{P_{X},X}\langle
P,{\cal{Q}}|J^{\nu}(0)|P_{X},X\rangle\langle
P_{X},X|\tilde{J}^{\mu}(q)|P,{\cal{Q}}\rangle\,,\nonumber\\
&=& 2\pi^{2}\sum_{X}\delta(M_{X}^{2}+[P+q]^{2})\langle P,
{\cal{Q}}|J^{\nu}(0)|P+q, X\rangle \langle
P+q,X|J^{\mu}(0)|P,{\cal{Q}}\rangle\,.
\end{eqnarray}
The momenta $P_{\mu}$, $q_{\mu}$, the polarization $n_{\mu}$ and the
currents $J_{\mu}$ are considered as four-dimensional quantities,
being their Lorentz indices raised and contracted with
$\eta^{\mu\nu}$.

In the large $N$ limit of the gauge theory only single hadron states
will contribute. In the gauge theory for the $s$-channel we have
\begin{eqnarray}\label{pol20}
s=-(P+q)^{2}\simeq q^{2}\bigg(\frac{1}{x}-1 \bigg)\,,
\end{eqnarray}
where it has been used $-P^{2}\ll q^{2}$. The corresponding
ten-dimensional scale is
\begin{eqnarray}\label{pol21}
\tilde{s} = -g^{MN}P_{X,M}P_{X,N}\leq
-g^{\mu\nu}(P+q)_{\mu}(P+q)_{\nu}
\sim\frac{R^{2}}{r^{2}_{int}}q^{2}\bigg(\frac{1}{x}-1
\bigg)=\frac{\bigg(\frac{1}{x}-1 \bigg)}{\alpha'(4\pi g_s
N)^{1/2}}\,.
\end{eqnarray}
The 't Hooft parameter appears in the denominator, so if $(g_s
N)^{-1/2}\ll x<1$ we have $\alpha'\tilde{s}\ll1$. Therefore, for
$x\sim1$ (with $x<1$) only massless string states are produced, and
we are dealing with a supergravity process \cite{Polchinski:2002jw}.
This is the regime studied in the present paper.

Now, let us focus on the way that the DIS process is viewed from the
bulk theory. Recall that in the four-dimensional boundary theory we
have to obtain the two-point function of two electromagnetic
currents inside the hadron. Using the gauge/string duality the
current operator inserted at the boundary of the $AdS$ space induces
a perturbation on the boundary condition of a bulk gauge field. The
perturbation excites a non-normalizable mode which propagates within
the bulk \cite{Gubser:1998bc,Witten:1998qj}. We can see how bulk
gauge fields appear as follows. The isometry group of the manifold
$W$ corresponds to an R-symmetry group on the field theory. We can
consider a $U(1)_R$ subgroup and the associated R-symmetry current
can now be identified with the electromagnetic current inside the
hadron. In addition, for the global symmetry group corresponding to
the isometry of manifold $W$, there is a Killing vector
$\upsilon_{j}$. It excites a non-normalizable mode of a Kaluza-Klein
gauge field,
\begin{eqnarray}\label{pol5}
\delta g_{mj}=A_{m}(y,r)\upsilon_{j}(\Omega)\,.
\end{eqnarray}
In the case of glueballs, the holographic dual field in
\cite{Polchinski:2002jw} corresponds to the dilaton. Thus, the
incoming bulk dilaton field $\Phi_i$ couples to the bulk
$U(1)$-gauge field $A_\mu$ (induced by a current inserted at the
boundary) and to another dilaton $\Phi_X$, which represents an
intermediate hadronic state (See Figure 1). The latter propagates in
the bulk and couples to an outgoing dilaton $\Phi_f$ (corresponding
to the final hadronic state) and a gauge field $A_\nu$ in the bulk
which comes from the insertion of a second boundary theory current.
This Witten diagram represents the holographic dual version of the
optical theorem \cite{Polchinski:2002jw} for a DIS process which is
schematically shown at the boundary slice, where the leptonic
currents are indicated with $l$, and they couple to the virtual
photons indicated by dashed lines at the boundary slice. Then, these
photons couple to the currents inside the hadrons, which are
indicated by solid lines on the boundary slice and are labeled by
the momenta carried by the initial, intermediate and final hadronic
states, $P_i$, $P_X$ and $P_f$, respectively.
\begin{figure}
\begin{center}
\epsfig{file=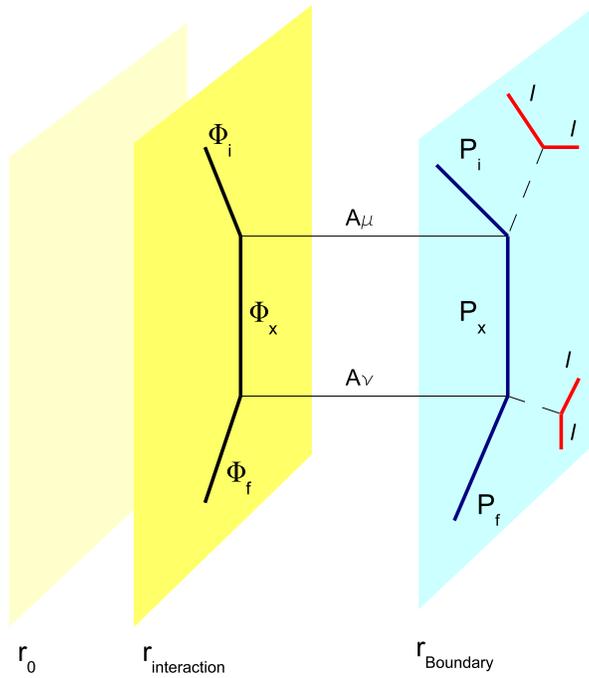, width=9cm}{\caption{\small Schematic
representation of the AdS/CFT prescription for the calculation of
the tensor $T_{\mu\nu}$. The interaction in the bulk occurs at
$r_{interaction}$, we also show the slices corresponding to the IR
cutoff $r_0$ and the boundary.}} \label{Figur1}
\end{center}
\end{figure}
The present holographic approach corresponds to a DIS process where
the lepton is scattered from an entire hadron, which becomes excited
but does not fragment. In the next two sections we extend these
ideas to other holographic dual models.

\section{DIS from mesons in the $\textmd{D3-D7}$-brane model}

In this section we firstly give a brief introduction to the
$\textmd{D3-D7}$-brane model developed in reference
\cite{Kruczenski:2003be}. In \cite{Karch:2002sh} it was shown that
by introducing $N_{f}$ D7-branes in the $AdS_{5}\times S^{5}$
background of type IIB string theory, it is possible to describe
$N_{f}$ hypermultiplets in the gauge theory preserving ${\cal{N}}=2$
supersymmetries in four dimensions. The starting point is a set of
coincident parallel $N$ D3-branes and $N_{f}$ D7-branes which share
directions 0-3 with the former ones. The hypermultiplets in the
gauge theory arise from the lightest modes of the fundamental
strings extended between D3 and D7-branes: those are modes of the
type 3-7 and 7-3, whose mass is $m_{q}=L/2\pi \alpha'$, where $L$ is
the distance between D3 and D7-branes in the (8, 9) plane. In the
decoupling limit of the D3-branes, namely $g_{s}N\gg1$ where $g_s$
is the string coupling, the background becomes $AdS_{5}\times
S^{5}$. Also, if $N\gg N_{f}$ then the backreaction of the D7-branes
can be neglected. In these limits the description corresponds to
$N_{f}$ probe D7-branes in the referred $AdS_{5}\times S^{5}$
geometry. We consider only the Abelian case ($N_{f}=1$), which
implies the existence of a global $U(1)$ symmetry group, associated
with the flavour symmetry of the ${\cal{N}}=2$ $SU(N)$ SYM theory.
This theory has dynamical quarks, thus it is very interesting to
investigate the DIS structure of scalar and vector mesons which
arise from this model. We should be aware that this model does not
allow to describe spontaneous breaking of chiral symmetry. This
makes a difference with respect to QCD and the Sakai-Sugimoto model
\cite{Sakai:2004cn}. In \cite{Kruczenski:2003be} the mass spectra
for different types of mesons have been computed through its
holographic dual description. In this work we shall obtain the
structure functions of the hadronic tensor for DIS from those scalar
and polarized vector mesons.

We develop a systematic method to derive the interaction Lagrangian
${\cal{L}}_{int}$ in the holographic dual theory, in a complementary
way with respect to the proposal of Polchinski and Strassler
\cite{Polchinski:2002jw}. We emphasize that the interaction
Lagrangian obtained using both methods is the same for the case
studied by Polchinski and Strassler \cite{Polchinski:2002jw}, while
it gives very different results for vector mesons in comparison with
\cite{Bayona:2009qe} and \cite{BallonBayona:2010ae}. The present
approach guarantees the existence of a conserved current in eight
dimensions straightforwardly derived from the Dirac-Born-Infeld
action. From this current we shall see that the property $q_{\mu}
W^{\mu\nu}=0$ is satisfied. Moreover, we have checked that
${\cal{L}}_{int}$ is invariant under parity and time-reversal
transformations. Thus, the hadronic tensor automatically satisfies
the properties indicated in Eqs.(\ref{DIS11})-(\ref{DIS15b}). All
these are consistency checks that confirm the correctness of the
procedure we develop here. Next step will be the derivation of the
structure functions for scalar and polarized vector mesons,
obtaining analytical expressions for all of them.

\subsection*{The $\textmd{D3-D7}$-brane model}

For this brief subsection we follow reference
\cite{Kruczenski:2003be}. Let us consider a type IIB string theory
background consisting of $N$ coincident D3-branes. This corresponds
to the geometry $AdS_{5} \times S^{5}$ whose metric is
\begin{eqnarray}\label{kru3}
ds^{2}=\frac{r^2}{R^{2}}\:ds^{2}({E}^{(1,3)})+\frac{R^{2}}{r^{2}} d\vec{Z}\cdot d\vec{Z} \,,
\end{eqnarray}
where the coordinates $Z^i$, $i=1, \ldots, 6$ parameterize the
transversal directions to the D3-branes and $r=\big|\vec{Z}\big|$.
In addition, the $AdS_5$ radius is given by $R^2=\sqrt{4\pi g_s N}
\alpha'$.

Let us add a probe D7-brane at a certain distance $|\vec{Z}|=L$ from
the D3-branes in the (8, 9) plane. In this case the hypermultiplet
becomes massive, and the R-symmetry is $SU(2)_{R}$. The induced
metric on the D7-brane is
\begin{eqnarray}\label{kru3}
ds^{2}=\frac{\rho^{2}+L^{2}}{R^{2}}\:ds^{2}({E}^{(1,3)})+\frac{R^{2}}{\rho^{2}+L^{2}}d\rho^{2}+
\frac{R^{2}\rho^{2}}{\rho^{2}+L^{2}}d\Omega_{3}^{2}\,,
\end{eqnarray}
where $\rho^{2}=r^{2}-L^{2}$ and $\Omega_{3}$ are the spherical
coordinates in the space spanned by the coordinates 4-7. For $L=0$
the metric reduces to $AdS_{5} \times S^{3}$, rendering a conformal
gauge theory. On the other hand, if $L\neq0$, the above metric is
only asymptotically $AdS_{5} \times S^{3}$ for $\rho \gg L$. This
shows the explicit breaking of conformal invariance induced by the
hypermultiplet mass $m_{q}= L/2\pi\alpha'$, which is recovered in
the high energy limit, $E\gg m_{q}$. Notice that the radius of the
$S^{3}$ is not a constant, and moreover it goes to zero for $\rho=0$
(which corresponds to $r=L$), where the D7-brane ends as seen from
the projection on the $AdS_{5}$.

In \cite{Kruczenski:2003be} the spectra of scalar and vector mesons
were computed, and they were arranged in supersymmetric multiplets
which transform according to representations of the global
$SU(2)_{R}\times SU(2)_{L}$ symmetry group. These mesons correspond
to excitations of open strings of the D7-brane. The dynamics of the
probe D7-brane fluctuations is described in terms of the following
action
\begin{eqnarray}\label{kru4}
S_{D7}&=& -\mu_{7}\int
d^{8}\xi\sqrt{-\textmd{det}(P[g]_{ab}+2\pi\alpha'F_{ab})}
+\frac{(2\pi\alpha')^{2}}{2}\mu_{7}\int P[C^{(4)}]\wedge F\wedge
F\,,
\end{eqnarray}
$g_{ab}$ stands for the metric (\ref{kru3}),
$\mu_{7}=[(2\pi)^{7}g_{s}\alpha'^{4}]^{-1}$ is the D7-brane tension
and $P$ denotes the pullback of the background fields on the
D7-brane world-volume. The relevant part of the Ramond-Ramond
potential in the Wess-Zumino term is
\begin{eqnarray}\label{kru5}
C^{(4)}=\frac{r^{4}}{R^{4}}dy^{0}\wedge dy^{1}\wedge dy^{2} \wedge
dy^{3}\, .
\end{eqnarray}

Next, in order to fix our notation and give a self-contained
derivation of the hadronic tensor, we show briefly how the mesons
are obtained from the present model. Then, we carry out the
derivation of the hadronic tensor and the structure functions for
scalar and vector mesons.

\subsection*{DIS from scalar mesons}

The equations of motion for scalar mesons are obtained from
transversal fluctuations of the D7-brane \cite{Kruczenski:2003be}
\begin{equation}\label{kru6a}
Z^{5}=0+2\pi \alpha'\chi\,,
\:\:\:\:\:\:\:\:\:\:\:\:\:\:\:Z^{6}=L+2\pi\alpha'\varphi\,,
\end{equation}
where the coordinates $Z^{5}$ and $Z^{6}$ lie on the (8, 9) plane,
transversal to the D7-brane. $\chi$ and $\varphi$ are the scalar
fluctuations whose Lagrangian is straightforwardly derived from
action (\ref{kru4}),
\begin{equation}\label{kru6}
{\mathcal{L}}=-\mu_{7}\sqrt{|{\det}(g_{ab})|}\sqrt{\det(\delta^{ab}+(g^{ac}R^{2}/r^{2})(2\pi\alpha')^{2}
(\partial_{c}\chi \partial^{b}\chi+\partial_{c}\varphi \partial^{b}
\varphi))}\,.
\end{equation}
All indices denote directions in the world-volume of the D7-brane.
Expanding the above Lagrangian up to quadratic order in the
derivatives of the fluctuations, the previous equation becomes
\begin{eqnarray}\label{kru7}
{\mathcal{L}}=-\mu_{7}\sqrt{|\det(g_{ab})|}
\bigg[1+\frac{2R^{2}}{r^{2}}g^{ab}(\pi\alpha')^{2}(\partial_{b}\chi
\partial_{a}\chi+\partial_{b}\varphi \partial_{a} \varphi)\bigg]\,,
\end{eqnarray}
which only depends upon derivatives of the scalars. The D7-brane
wraps the $S^3$, whose radius is denoted by $\rho$. Substituting
$r^{2}=\rho^{2}+L^{2}$ and the metric (\ref{kru3}) in the quadratic
Lagrangian one obtains the equations of motion (EOM) for scalar
fluctuations of the D7-brane in the probe approximation, which reads
\begin{equation}\label{kru8}
\partial_{a} \bigg(\frac{\rho^{3}\sqrt{\textmd{det}\tilde{g}}}{\rho^{2}+L^{2}}
g^{ab}\partial_{b}\Phi \bigg)=0\,,
\end{equation}
where $\Phi$ is either fluctuation ($\chi$ or $\varphi$), while
$\tilde{g}_{ij}$ is the metric of a 3-sphere of unit radius, which
together with $\rho$ include the directions ($Z^{1}$, $\cdots$,
$Z^{4}$). The EOM can be written more explicitly as
\begin{eqnarray}\label{kru9}
\frac{R^{4}\rho}{(\rho^{2}+L^{2})}\delta^{\mu\nu}\partial_{\mu}\partial_{\nu}\Phi
+\partial_{\rho}\bigg(\frac{\rho^{5}}{(\rho^{2}+L^{2})}\partial_{\rho}\Phi
\bigg)
+\frac{\rho^{3}}{(\rho^{2}+L^{2})}\nabla^{i}\nabla_{i}\Phi=0\,,
\end{eqnarray}
where $\nabla_{i}$ is the covariant derivative on $S^{3}$. Indices
$a$, $b$, $c$, $\cdots$ denote coordinates on the D7-brane
world-volume; $i$, $j$, $k$, $\cdots$ are on $S^{3}$;  $\mu$, $\nu$,
$\cdots$ indicate directions parallel to the D3-branes and $m$, $n$,
$\cdots$ correspond to $0, 1, 2, 3, \rho$ coordinates.

Now, we need the wave functions for the scalar mesons in a region
where
\begin{equation}\label{kru10}
r_{int}=\sqrt{\rho_{int}^{2}+L^2}\sim q R^{2}\gg r_{0} \equiv
\Lambda R^{2} > \sqrt{2}L\:\:\Rightarrow\:\:\rho_{int}\gg L\, ,
\end{equation}
where $r_{int}$ denotes the interaction region and $\Lambda$ is an
infrared cutoff for the four-dimensional gauge theory. In this
region the metric can be approximated by $AdS_{5}\times S^{3}$,
which reflects the fact that for energies as high as $E\gg m_{q}$
the explicitly broken conformal invariance gets restored. Thus, the
resulting EOM is
\begin{eqnarray}\label{kru11}
\frac{R^{4}}{\rho}\delta^{\mu\nu}\partial_{\mu}\partial_{\nu}\Phi
+\partial_{\rho}(\rho^{3}\partial_{\rho}\Phi)
+\rho\nabla^{i}\nabla_{i}\Phi=0\,.
\end{eqnarray}
The proposed Ansatz for the solution is
\begin{equation}\label{kru12}
\Phi^{\ell}=\phi(\rho)\:e^{iP \cdot y}\:Y^{\ell}(S^{3})\,,
\end{equation}
where $Y^{\ell}(S^{3})$ are the scalar spherical harmonics on
$S^{3}$, which satisfy the eigenvalue equation
\begin{equation}\label{kru13}
\nabla^{i}\nabla_{i}Y^{\ell}(S^{3})=-\ell(\ell+2)Y^{\ell}(S^{3})\,.
\end{equation}
Notice that in order to simplify the notation, from now on we will
call $\Phi \equiv \Phi^{\ell}$, omitting the $\ell$ index.

Thus, the Lagrangian (\ref{kru7}) for the complex scalar $\Phi$ in
the interaction region can be written as
\begin{equation}\label{kru14}
{\mathcal{L}}=-\mu_{7}(\pi \alpha')^{2}\sqrt{|\textmd{det}
g|}\frac{R^{2}}{\rho^{2}}g^{ab}(\partial_{a}\Phi
\partial_{b}\Phi^{*}+\partial_{b}\Phi \partial_{a}\Phi^{*})\,.
\end{equation}
Now, we have to derive the EOM for the fields in the $AdS$ space.
Thus, replacing the solution (\ref{kru12}) for the scalar mesons in
the interaction region defined by Eq.(\ref{kru10}), we obtain
\begin{equation}\label{kru19}
\partial_{\rho}^{2}\phi+3\rho^{-1}\partial_{\rho}\phi+(-P^{2}R^{4}\rho^{-4}-m^{2}_{\ell}\rho^{-2})\phi=0\,,
\end{equation}
with $m^{2}_{\ell}=\ell(\ell+2)$.

For the initial/final hadronic state (IN/OUT), with four-momentum
$P$, we can use the leading behaviour of the solution in the region
$\rho\sim\rho_{int}$, while for the intermediate state  $X$ with
four-momentum $P_{X}$, we have to use the full solution. Thus, we
obtain
\begin{equation}\label{kru20}
\Phi_{IN/OUT}=c_{i}(\Lambda \rho)^{-\nu-1}e^{iP \cdot y}Y(\Omega)\,,
\end{equation}
\begin{equation}\label{kru21}
\Phi_{X}=c_{X}\Lambda^{-3/2}s^{1/4}
\rho^{-1}J_{\nu}\bigg(\frac{s^{1/2}R^{2}}{\rho}\bigg)e^{iP_{X} \cdot
y}Y(\Omega)\,,
\end{equation}
where $J_{\nu}$ is the Bessel function of first kind,
$\nu=\sqrt{m^{2}_{\ell}+1}=\ell+1$ and $s=-(P+q)^2=M_X^2$ is the
mass-squared of the intermediate state, while $c_{X}$ and $c_{i}$
are dimensionless constants. In order to study the parity we need to
use the property
$[Y^{\ell}(\Omega)]_{\textmd{P}}=(-1)^{\ell}Y^{\ell}(\Omega)$. As a
result, the mesons are scalar and pseudoscalar mesons for even and
odd values of $\ell$, respectively.

At this point we have all the ingredients to calculate the hadronic
tensor using the holographic dual prescription. For this purpose we
first consider that the holographic meson couples to a gauge field
in the bulk of the string theory dual model. In order to understand
this coupling let us consider the insertion of a $U(1)$-current at
the $AdS_5$ boundary. This induces a perturbation on the boundary
conditions and excites a non-normalizable mode $A_{m}$ in the
$AdS_5$ interior. This mode corresponds to the extension towards the
$AdS$-interior of the $A_{\mu}|_{4d}$ field, which is in fact its
boundary value. The perturbation propagates inside the $AdS$,
inducing a metric fluctuation which is the product of the gauge
field $A_{m}(y,\rho)$ and a Killing vector $\upsilon_{j}$
corresponding to an isometry of the internal space $S^3$. Thus, we
consider the fluctuation proposed in \cite{Polchinski:2002jw}.
Perturbing the metric in Eq.(\ref{kru14}) we have $g^{ab}\rightarrow
g^{ab}-\delta g^{ab}$, where $\delta g^{ab}=A^{m}(y,
\rho)\upsilon^{j}(\Omega)$ and using
$\upsilon^{j}\partial_{j}Y^\ell(\Omega)=i{\cal{Q}}_\ell
Y^\ell(\Omega)$ (below we shall drop the $\ell$ label from the
spherical harmonics and from ${\cal{Q}}_\ell$), the interaction
Lagrangian we obtain is:
\begin{equation}\label{kru15}
{\mathcal{L}}_{int}=i{\cal{Q}}\mu_{7}(\pi
\alpha')^{2}\sqrt{|\textmd{det} g|}R^{2}\rho^{-2}A^{m}(\Phi
\partial_{m}\Phi^{*}_{X}-\Phi^{*}_{X}\partial_{m}\Phi )\,.
\end{equation}
The angular dependence on the spherical harmonics corresponds to
functions which are charge eigentates, with charge ${\cal{Q}}$ under
the $U(1)$ symmetry group induced by transformations on the internal
$S^3$ in the direction of the Killing vector $\upsilon^{j}$.

Alternatively, ${\cal{L}}_{int}$ can be obtained from the coupling
of the gauge field $A_{m}$ to the Noether current corresponding to
the internal symmetry of the action (\ref{kru14}) under global phase
transformations associated with the $U(1)$ group, which is a
subgroup of the $SO(4)$ isometry group of $S^3$, generated by a
Killing vector $\upsilon^{j}$. The Noether's current is:
\begin{equation}\label{kru16}
j^{m}=\frac{\partial{\cal{L}}}{\partial(\partial_{m}\Phi)}\Theta
+\frac{\partial{\cal{L}}}{\partial(\partial_{m}\Phi^{*})}\Theta^{*}\,.
\end{equation}
Now, for an infinitesimal transformation with parameter
$\vartheta/2(=\textmd{constant})$ we have
\begin{eqnarray}\label{kru17}
\delta\Phi=-i\frac{\vartheta}{2}\Phi\:\:\Rightarrow \:\:\Theta=-\frac{i}{2}\Phi\,,\nonumber\\
\delta\Phi^{*}=i\frac{\vartheta}{2}\Phi^{*}\:\:\Rightarrow
\:\:\Theta^{*}=\frac{i}{2}\Phi^{*}\,.
\end{eqnarray}
Therefore, the Noether's current reads
\begin{equation}\label{kru18}
j^{m}=i\mu_{7}(\pi \alpha')^{2}R^{2}\rho^{-2}(\Phi
\partial^{m}\Phi^{*}_{X}-\Phi^{*}_{X}\partial^{m}\Phi )\,.
\end{equation}
If we define ${\cal{L}}_{int}={\cal{Q}}\sqrt{|\textmd{det}
g|}A^{m}j_{m}$, we obtain the same ${\cal{L}}_{int}$ given by
Eq.(\ref{kru15}) from a metric fluctuation. Consequently,
${\cal{L}}_{int}$ is given by the coupling of the gauge field
$A^{m}$ to the conserved Noether's current $j_{m}$. Now, a crucial
point for the present derivation of the hadronic tensor is that the
fields (\ref{kru12}) are charged under a $U(1)$ symmetry group,
which is a subgroup of the $SO(4)$ isometry group of $S^3$,
generated by a Killing vector $\upsilon^{j}$. In order to check this
we may ask how they transform under a coordinate transformation
corresponding to an isometry on $S^{3}$: $(x^{m},
\,x^{i})\rightarrow (x^{m}, \,x^{i}+\upsilon^{i} \epsilon(x^{m}))$,
where $\upsilon^{i}$ is the Killing vector which generates the
isometry. We consider $\ell>0$ which means that the scalar fields
are charged under a $U(1)$ group.

In order to calculate the matrix element relevant for the hadronic
tensor, we will use the identification \cite{Polchinski:2002jw}
\begin{equation}\label{kru21.1}
S_{int}=(2\pi)^4 \delta^4(P_X-P-q)n_{\mu}\langle P+q, X|J^{\mu}(0)|P, {\cal {Q}}\rangle \,.
\end{equation}
Therefore, we must calculate $S_{int}$. We start by deriving the
coupling $A_{m}j^{m}$, using the expressions
\begin{eqnarray}
A_{\mu}&=& n_{\mu}e^{iq \cdot y}\frac{qR^{2}}{\rho}K_{1}\bigg(\frac{q R^{2}}{\rho}\bigg)\,,\nonumber \\
A_{\rho}&=&
-\frac{i}{q^{2}}\eta^{\mu\nu}q_{\mu}\partial_{\rho}A_{\nu}\,,
\end{eqnarray}
where $A_{\mu}$ and $A_{\rho}$ are the components of $A_{m}(y,
\rho)$ deduced in \cite{Polchinski:2002jw}. The current conservation
equation is
$\partial_{\nu}j^{\nu}+\rho^{-3}\partial_{\rho}(\rho^{3}j^{\rho})=0$.
We obtain
\begin{equation}\label{kru100}
A_{m}j^{m}=A_{\mu}\bigg(j^{\mu}-i
\frac{q^{\mu}}{q^{2}}\partial_{\gamma}j^{\gamma}\bigg)-i\frac
{q^{\mu}}{q^{2}\rho^{3}}\partial_{\rho}(A_{\mu}\rho^{3}j^{\rho})\,.
\end{equation}
Then, the interaction is given by
\begin{eqnarray}\label{kru101}
S_{int} &=& \int_{r_{0}}^{\infty} d^{8}x{\cal{Q}}\sqrt{|\textmd{det} g|}A_{m}j^{m} \nonumber\\
&=& \int_{r_{0}}^{\infty} d^{8}x 2\mu_{7}
(\pi\alpha')^{2}{\cal{Q}}\sqrt{|\textmd{det}
\tilde{g}|}\rho^{-1}R^{4}\bigg(P^{\mu}+\frac{q^{\mu}}{2x}\bigg)A_{\mu}
\Phi\Phi_{X}^{*} \nonumber \\
&& -i \frac {q^{\mu}}{q^{2}R^{3}}\int d^{4}y
\:d\Omega {\cal{Q}}\sqrt{|\textmd{det}
\tilde{g}|}(A_{\mu}\rho^{3}j^{\rho})|^{\infty}_{\Lambda R^{2}}\,.
\end{eqnarray}
The integral can be calculated with the substitution $z\equiv
R^{2}/\rho$ and taking $1/\Lambda\rightarrow\infty$ since $A_\mu$
goes to zero very fast for $\rho < \rho_{int}$. We obtain that the
second term of Eq.(\ref{kru101}) evaluated at the integration limits
vanishes. After integration of the remaining term, and using
Eq.(\ref{kru21.1}), it results
\begin{equation}\label{kru102}
\langle P+q, X|J^{\mu}(0)|P, {\cal {Q}}\rangle \,= \kappa_1
I_1\bigg(P^{\mu}+\frac{q^{\mu}}{2x}\bigg)\,,
\end{equation}
where
$\kappa_1=2\mu_{7}{\cal{Q}}(\pi\alpha')^{2}\:c_{i}c_{X}^{*}\Lambda^{\nu-5/2}q^{3/2}
x^{-1/4} (1-x)^{1/4}$ and the result of the integral in $\rho$ is
$I_1$. For $|t|\ll 1$ we can approximate $s\simeq q^{2}(1/x-1)$.
Doing this, we obtain
\begin{equation}
I_1=2^{1+\nu}\Gamma(2+\nu)q^{-3-\nu}(1-x)^{\nu/2}x^{\nu/2+2}\,.
\end{equation}
In order to compute $\textmd{Im}\:T^{\mu\beta}$ as in
\cite{Polchinski:2002jw}, we have to multiply Eq.(\ref{kru102}) by
its complex conjugate and sum over radial excitations. For this we
have to know the density of states, which can be estimated by
introducing an IR cutoff at $r_{0}$. In this way the distance
between zeros of the Bessel function of Eq.(\ref{kru21}) gives
$M_{n}=n\pi\Lambda$. In the large $N$ limit this becomes a sum of
delta functions, and for large $q$ we have
\begin{eqnarray}\label{kru103}
\sum_{n}\delta(M_{n}^{2}-s)\sim \bigg(\frac{\partial
M_{n}^{2}}{\partial n}\bigg)^{-1}\sim (2\pi s^{1/2}\Lambda)^{-1}\,.
\end{eqnarray}
Assembling all these partial results, we obtain
\begin{eqnarray}\label{kru22}
\textmd{Im} \:T^{\mu\beta} &=& 4\pi^{5}{\cal{Q}}^{2}2^{2+2\nu}
[\Gamma(2+\nu)]^{2}\mu_{7}^{2}(\alpha')^{4}|c_{i}|^{2}|c_{X}|^{2}
\nonumber\\ & &  \times \Lambda^{2\nu-6}
q^{-4-2\nu}(1-x)^{\nu}x^{\nu+4}\bigg(P^{\mu}+\frac{q^{\mu}}{2x}\bigg)
\bigg(P^{\beta}+\frac{q^{\beta}}{2x}\bigg)\,. \,\,\,\,\,\,\,\,\: \,\,\,\,\,\,\,\,\:
\end{eqnarray}
The remaining step is to obtain the structure functions for the
scalar mesons from Eq.(\ref{DIS51}). It can be seen that
$W^{\mu\beta}$ satisfies the properties described above. Then, the
structure functions we obtain are:
\begin{equation}\label{kru23}
F_{1}=0,
\,\,\,\,\,\,\,\,\:\,\,\,\,\,\,\,\,\:\:\:\:F_{2}=A'_{0}{\cal{Q}}^{2}
\bigg(\frac{\mu_{7}^{2}\alpha'^{4}}{\Lambda^{8}}\bigg)
\bigg(\frac{\Lambda^{2}}{q^{2}}\bigg)^{\nu+1}x^{\nu+3}(1-x)^{\nu}\,,
\end{equation}
where
$A'_{0}=4\pi^{5}2^{2+2\nu}|c_{i}|^{2}|c_{X}|^{2}[\Gamma(2+\nu)]^{2}$
is a dimensionless normalization constant. We shall comment about
these results in the last section of the paper.

\subsection*{DIS from vector mesons}

Now, we focus on how to derive the structure functions for polarized
vector mesons in the D3-D7-brane model. First we derive
${\cal{L}}_{int}$ for the coupling of a gauge field $A_{m}$ to
vector mesons. For this purpose we must find explicit expressions
for the fields. With this we obtain all the structure functions for
polarized vector mesons, which have not been considered in the
previous literature in the context of the gauge/string duality.

In \cite{Kruczenski:2003be} the spectrum of vector mesons has been
computed. They arise from fluctuations of the vector fields of the
Dirac-Born-Infeld (DBI) action of the probe D7-brane, in the
directions parallel to this brane. The starting point is the action
(\ref{kru4}). As before, first the EOM are derived from it and then
the Lagrangian is expanded up to quadratic order. This new
Lagrangian gives the same EOM as (\ref{kru4}). Recall that in
(\ref{kru4}) $F^{ab}=\partial^{a}B^{b}-\partial^{b}B^{a}$. The EOM
is
\begin{eqnarray}\label{kru26}
\partial_{a}(\sqrt{|\textmd{det} g|}F^{ab})-\frac{4}{R^{4}}\rho(\rho^{2}+L^{2})
\varepsilon^{bjk}\partial_{j}B_{k}=0\,,
\end{eqnarray}
where $\varepsilon^{ijk}$ is the Levi-Civita pseudo-tensor, the
indices $a$, $b$, $c$, $d$, $\cdots$ run over all directions of the
D7-brane world-volume, and $i$, $j$, $k$, $\cdots$ belong to
$S^{3}$. Also, the second term, coming from the Wess-Zumino
Lagrangian, is only present if $b$ is on $S^{3}$. The Ansatz
\cite{Kruczenski:2003be} for the solution of the vector mesons
$B_{\mu}$ is:
\begin{eqnarray}\label{kru27}
B_{\mu}=\zeta_{\mu}\:\phi(\rho)\:e^{iP \cdot
y}\:Y^{\ell}(S^{3}),\:\:\:\:\:P \cdot
\zeta=0,\:\:\:\:\:B_{\rho}=0,\:\:\:\:\:B_{i}=0\,,
\end{eqnarray}
where there has been done an expansion in spherical harmonics on
$S^{3}$, $\phi(\rho)$ is a function to be determined, $\zeta_{\mu}$
is the polarization vector and the relation $\zeta \cdot P=0$ comes
from $\partial^{\mu}B_{\mu}=0$.

The interaction region corresponds to $\rho_{int}\gg L$ with
$L\neq0$, where the energy in four dimensions specified in
(\ref{pol2}) is $E^{(4)}\gg \Lambda$. As in the case of the scalar
mesons above, in this regime the metric can be approximated by a
pure $AdS_{5}\times S^{3}$ metric. Now, in the interaction region of
the $AdS$ space the EOM for $b = \mu$ is given by
\begin{eqnarray}\label{kru33}
\frac{1}{\sqrt{|\textmd{det}g_{AdS}|}}\partial_{m}
(\sqrt{|\textmd{det}g_{AdS}|}\partial^{m}B^{\mu})-\frac{m^{2}_{\ell}}{R^{2}}B^{\mu}=0\,,
\end{eqnarray}
where $g_{AdS}$ is the $AdS_{5}$ metric, $m$, $n$, $\cdots$, denote
$0123\rho$ coordinates and $m^{2}_{\ell}=\ell(\ell+2)$. Replacing
the proposed Ansatz (\ref{kru27}) in Eq.(\ref{kru33}) gives the same
equation as Eq. (\ref{kru19}), with the solution
\begin{eqnarray}\label{kru35}
\phi(\rho)=\frac{c}{\rho}J_{\nu}\bigg(\frac{s^{1/2}R^{2}}{\rho}\bigg)\,,
\end{eqnarray}
where $J_{\nu}$ is the Bessel function of first kind, $c$ is a
dimensionless constant and $\nu=\ell+1$. For the initial and final
hadronic states we can consider only the leading behaviour for
$|P^{2}|\ll q^{2}$, so the solution is
\begin{eqnarray}\label{kru36}
B_{\mu \,IN/OUT}=\zeta_{\mu}c_{i}\Lambda^{-1}(\Lambda
\rho)^{-\nu-1}e^{iP \cdot y}\:Y^{\ell}(S^{3})\,,
\end{eqnarray}
where $c_{i}$ is a dimensionless normalization
constant. For the intermediate state we consider the full solution
\begin{eqnarray}\label{kru37}
B_{X \mu}=\zeta_{X \mu}c_{X}\frac{s^{-1/4}\Lambda^{-3/2}}{\rho}
J_{\nu}\bigg(\frac{s^{1/2}R^{2}}{\rho}\bigg)e^{iP_{X} \cdot
y}\:Y^{\ell}(S^{3})\,,
\end{eqnarray}
where $c_{X}$ is a dimensionless normalization constant. In this
case the radius of the Bessel function $s^{1/2}R^{2}$ is much larger
than $r_{0}=\Lambda R^2$, implying that the full solution must be
used. To study the parity transformations, using the relations
$[Y^{\ell}(\Omega)]_{\textmd{P}}=(-1)^{\ell}Y^{\ell}(\Omega)$ and
$[\zeta_{\mu}]_{\textmd{P}}=(\zeta^{0}, -\vec{\zeta})$ we can
classify the mesons as vector mesons for even values of $\ell$ and
axial vector mesons for odd values of $\ell$.

From the expansion of $B_{\mu}$ in spherical harmonics on $S^{3}$,
it can be seen that the gauge fields on the D7-brane correspond to
charged fields in $AdS_{5}$. To see this we can write the full
Kaluza-Klein expansion
\begin{eqnarray}\label{kru38}
\breve{B}_{\mu}=\sum_{\ell}\,a_{\ell}\,B_{\mu}^{\ell}\,,
\end{eqnarray}
where $a_{\ell}$ are coefficients. Then, we consider coordinate
transformations on the higher-dimensional spacetime such as
$(x^{m},x^{i})\rightarrow (x^{m},x^{i}+\upsilon^{i}
\varepsilon(x^{m}))$, where $\upsilon^{i}$ is a Killing vector on
$S^{3}$, $i$ denotes $S^{3}$ coordinates and $m$ belongs to
$AdS_{5}$. For the variation of the vector field under this
transformation (being $\breve{B}_{\mu}$ a gauge field on
$AdS_{5}\times S^{3}$) we find
\begin{eqnarray}\label{kru39}
\delta
\breve{B}_{\mu}=-\epsilon(x^{m})\upsilon^{i}\partial_{i}\breve{B}_{\mu}-\partial_{\mu}\epsilon(x^{m})\,,
\end{eqnarray}
using the eigenvalue equation
$\upsilon^{j}\partial_{j}Y^{\ell}(S^{3})=i{\cal{Q}}_\ell
Y^{\ell}(S^{3})$,
\begin{eqnarray}\label{kru40}
\delta B_{\mu}^{0} &=& -\partial_{\mu}\epsilon(x^{m})
\:\:\:\:\:\:\:\:\:\:\:\:\:\:\:\:\:\:\:\:\:\:\:\:\:\ell=0\,,\\
\delta B_{\mu}^{\ell} &=& -\epsilon(x^{m})i{\cal{Q}}_\ell
B_{\mu}\:\:\:\:\:\:\:\:\:\:\:\:\:\:\:\:\:\ell>0\,.
\end{eqnarray}
Thus, we obtain a gauge field $B_{\mu}^{0}$ since it transforms
accordingly under the $U(1)$ transformations parameterized by
$\epsilon(x)$. On the other hand, all the fields with $\ell
> 0$ have a charge ${\cal{Q}}_\ell$, since they transform  with a phase under
the $U(1)$ induced by a transformation on $S^{3}$ in the direction
of the Killing vector. In addition, all the $B_{\mu}^{\ell}$ with
$\ell>0$ are massive vector fields from the $AdS_{5}$ point of view,
being their masses given by $m^{2}_{\ell}=\ell(\ell+2)$ (the
dimensional $R^2$ factor has been ignored), since they are not
invariant under gauge transformations. Their EOM is
Eq.(\ref{kru33})\footnote{We will use $B_{\mu}^{\ell}\equiv B_{\mu}$
as our notation, with $\ell>0$, therefore there is a field $B_{\mu}$
for each $\ell$.}. The EOM for the vector mesons in the interaction
region can also be derived from the following quadratic Lagrangian,
\begin{eqnarray}\label{kru41}
{\mathcal{L}}=-2\mu_{7}(\pi \alpha')^{2}\sqrt{|\textmd{det}
g|}F^{ab}F_{ab}^{*}\,.
\end{eqnarray}

Now, let us focus on the bulk interactions. Considering the metric
fluctuation $\delta g_{mj}=A_{m}(y,r)\upsilon_{j}(\Omega)$ and
$\upsilon^{j}\partial_{j}Y(\Omega)=i{\cal{Q}}Y(\Omega)$ we obtain
the following interaction Lagrangian,
\begin{equation}\label{kru42}
{\mathcal{L}}_{int}=2i{\cal{Q}}\mu_{7}(\pi
\alpha')^{2}\sqrt{|\textmd{det}
g|}A_{m}[B_{Xn}^{*}F^{nm}-B_{n}(F_{X}^{nm})^{*}]\,.
\end{equation}
which alternatively can be obtained from the coupling of the gauge
field $A_{m}$ to the Noether's current corresponding to the internal
symmetry of the action (\ref{kru41}). This corresponds to $U(1)$
transformations, thus ${\cal{L}}_{int}={\cal{Q}}\sqrt{|\textmd{det}
g|}A_{m}j^{m}$ where
\begin{equation}\label{kru43}
j^{m}=2i\mu_{7}(\pi
\alpha')^{2}[B_{Xn}^{*}F^{nm}-B_{n}(F^{nm}_{X})^{*}]\,.
\end{equation}
Therefore, ${\cal{L}}_{int}$ results from the coupling of the gauge
field  $A_{m}$ to the conserved current $j^{m}$. At this point, we
need to compute the coupling $A_{m}j^{m}$. We use a similar method
to that described before for scalar mesons, thus
\begin{eqnarray}\label{kru45}
S_{int}&=& \int_{r_{0}}^{\infty} d^{8}x{\cal{Q}}\sqrt{|\textmd{det} g|}A_{m}j^{m} \nonumber\\
&=& \int_{r_{0}}^{\infty} d^{8}x 2\mu_{7}
(\pi\alpha')^{2}{\cal{Q}}\sqrt{|\textmd{det} g|}R^{4}\rho^{-4}
A_{\mu}N^{\mu}\tilde{B}\tilde{B}_{X}^{*} -i \frac
{q^{\mu}}{q^{2}R^{3}}\int d^{4}y
d\Omega(A_{\mu}\rho^{3}j^{\rho})|^{\infty}_{\Lambda
R^{2}}\,,\,\,\,\,\,\,\,\,\,
\end{eqnarray}
where $B_{\mu}=\zeta_{\mu}\tilde{B}$ and
\begin{equation}\label{kru46a}
N^{\mu}=2(\zeta \cdot
\zeta_{X})\bigg(P^{\mu}+\frac{q^{\mu}}{2x}\bigg)+(\zeta_{X} \cdot
q)\zeta^{\mu}-(\zeta \cdot q)\zeta^{\mu}_{X}\,.
\end{equation}
We consider $z\equiv R^{2}/\rho$ and $1/\Lambda\rightarrow\infty$ in
order to calculate the integral analytically, for the same reason as
in the previous case. We obtain again that the second term of
Eq.(\ref{kru45}) evaluated at the integration limits vanishes.

Then, comparing with the Ansatz in Eq.(\ref{kru21.1}), we find
\begin{equation}\label{kru46}
\langle P+q, X|J^{\mu}(0)|P, {\cal {Q}}\rangle \,=\kappa_2 I_2
N^{\mu}\,,
\end{equation}
where
$\kappa_2=2\mu_{7}{\cal{Q}}(\pi\alpha')^{2}c_{i}c_{X}^{*}\,q^{1/2}\Lambda^{\nu-7/2}x^{1/4}(1-x)^{-1/4}$
and the integral in  $\rho$ is $I_2$, using $s\simeq q^{2}(1/x-1)$
this reduces to
\begin{equation}
I_2=2^{\nu+1}\Gamma(\nu+2)\,q^{-\nu-3}x^{\nu/2+2}(1-x)^{\nu/2}\,.
\end{equation}
In order to obtain $\textmd{Im}\:T^{\mu\nu}$ from Eq.(\ref{pol3}),
we have to multiply Eq.(\ref{kru46}) by its complex conjugate and
sum over the radial excitations and over the polarizations of the
final hadronic states $\zeta_{X}^{\mu}$. The density of states can
be estimated as for the scalar mesons. Then, the resulting
expression for the tensor is
\begin{eqnarray}\label{kru47}
\textmd{Im} \:T^{\mu\nu} &=& \frac{\pi\kappa_2
\kappa_2^{*}I_2^{2}}{\Lambda s^{1/2}}\sum_{\lambda}N^{\mu}N^{*\nu}\,,\nonumber\\
&=&\frac{\pi\kappa_2 \kappa_2^{*}I_2^{2}}{\Lambda
s^{1/2}}\sum_{\lambda}\bigg\{\bigg[2(\zeta \cdot
\zeta_{X})\bigg(P^{\mu}+\frac{q^{\mu}}{2x}\bigg)+(\zeta_{X} \cdot q)
\zeta^{\mu}-(\zeta \cdot q)\zeta^{\mu}_{X}\bigg]\nonumber\\ & &
\times \bigg[ 2(\zeta^{*} \cdot
\zeta^{*}_{X})\bigg(P^{\nu}+\frac{q^{\nu}}{2x}\bigg)+(\zeta^{*}_{X}\cdot
q) \zeta^{*\nu}-(\zeta^{*} \cdot q)\zeta^{*\nu}_{X}\bigg]\bigg\}\,.
\end{eqnarray}
For the solution (\ref{kru27}) of vector mesons, we have $P \cdot
\zeta=0$. Thus, there are three allowed polarizations: $\lambda=1, 2, 3$.
The normalization for the polarizations is $\zeta^{\mu}(P_{X},\lambda)
\cdot \zeta_{\mu}^{*}(P_{X},\lambda')=-M^{2}_{X}\delta_{\lambda,\lambda'}$ then
\begin{eqnarray}\label{kru48}
\sum_{\lambda}\zeta_{X\mu}(P_{X},\lambda)\zeta_{X\nu}^{*}(P_{X},\lambda)
&=& M^{2}_{X} \eta_{\mu\nu}+P_{X\mu}P_{X\nu} \nonumber\\
&=& M^{2}_{X}\eta_{\mu\nu}+(P_{\mu}+q_{\mu})(P_{\nu}+q_{\nu})\,.
\end{eqnarray}
We must sum over the polarizations of the final hadronic state $X$.
$\textmd{Im} T_{\mu\nu}$ can be written in terms of a symmetric and
an antisymmetric part, also we can neglect terms proportional to
$q_{\mu}$ and $q_{\nu}$, thus obtaining
\begin{equation}\label{kru49}
\sum_{\lambda}\textmd{Im} T_{\mu\nu}\equiv
H_{\mu\nu}=H_{\mu\nu}^{S}+H_{\mu\nu}^{A}\,,
\end{equation}
where $H_{\mu\nu}^{S}$ and $H_{\mu\nu}^{A}$ are the symmetric and
antisymmetric parts of $H_{\mu\nu}$ respectively. They result
\begin{eqnarray}\label{kru50}
H_{\mu\nu}^{S} &=&\frac{2\pi\kappa_2 \kappa_2^{*}I_2^{2}}{\Lambda
s^{1/2}}\bigg\{-\eta_{\mu\nu}(\zeta.q)(\zeta^{*}.q)(P+q)^{2}+P_{\mu}P_{\nu}\bigg[-4P^{2}(P+q)^{2}+(q.\zeta)(q.\zeta^{*})\bigg]
\nonumber\\ & & +(\zeta_{\mu}\zeta_{\nu}^{*}+
\zeta_{\nu}\zeta_{\mu}^{*})\frac{1}{2}\bigg[(P\cdot
q)^2-P^2q^2\bigg] \nonumber\\& & +(P_{\mu}\zeta^{*}_{\nu}+P_{\nu}
\zeta^{*}_{\mu})(\zeta\cdot q)\frac{1}{2}\bigg[P\cdot q+q^{2}\bigg]
+(P_{\mu}\zeta_{\nu}+P_{\nu}\zeta_{\mu})(\zeta^{*}\cdot
q)\frac{1}{2}\bigg[P\cdot q+q^{2}\bigg]\bigg\}\,,
\end{eqnarray}
and
\begin{eqnarray}\label{kru51}
H_{\mu\nu}^{A}&=& \frac{2\pi\kappa_2 \kappa_2^{*}I_2^{2}}{\Lambda
s^{1/2}}\bigg\{\frac{1}{2}(\zeta_{\mu}\zeta_{\nu}^{*}-\zeta_{\nu}\zeta_{\mu}^{*})\bigg[(P\cdot
q)^2-P^2q^2\bigg]\nonumber\\ & &
+(P_{\nu}\zeta^{*}_{\mu}-P_{\mu}\zeta_{\nu}^{*})(\zeta.q)\frac{1}{2}\bigg[4P^{2}+7P.q+3q^{2}\bigg]\nonumber\\
& & +
(P_{\mu}\zeta_{\nu}-P_{\nu}\zeta_{\mu})(\zeta^{*}.q)\frac{1}{2}\bigg[4P^{2}+7P.q+3q^{2}\bigg]
\bigg\}\,.
\end{eqnarray}
Now, we have to rewrite the hadronic tensor for spin-one hadrons
$W_{\mu\nu}$ from Eq.(\ref{DIS16}) in order to compare it with
$H_{\mu\nu}$.

Using Eq.(\ref{DIS51}) and comparing it with the hadronic tensor for
polarized targets of spin one, we obtain a system of 6 equations and
8 functions to determine, which are the structure functions.
However, since these functions do not depend on the polarizations,
there are two additional equations, and therefore the structure
functions become completely determined. The structure functions can
be written in terms of dimensionless variables
$t=\frac{P^{2}}{q^{2}}<0$ and $x=-\frac{q^{2}}{2P \cdot q}$, as
defined in Section \ref{Kinematics}, and also in terms of the model
dependent set of parameters such as $\Lambda$, $|c_i|$, $|c_X|$ and
${\cal {Q}}$. In this way, we obtain the following set of DIS
structure functions:
\begin{eqnarray}
F_{1}&=& A(x)\frac{1}{12x^{3}}(1-x-2xt-4x^{2}t+4x^{3}t+8x^{3}t^{2})\,,\\
F_{2}&=& A(x)\frac{1}{6x^{3}}(1-x+12xt-14x^{2}t-12x^{2}t^{2})\,,\\
b_{1}&=& A(x)\frac{1}{4x^{3}}(1-x-xt)\,,\\
b_{2}&=& A(x)\frac{1}{2x^{3}}(1-x-x^{2}t)\,,\\
b_{3}&=& A(x)\frac{1}{24x^{3}}(1-4x+8x^{2}t)\,,\\
b_{4}&=& A(x)\frac{1}{12x^{3}}(-1+4x-2x^{2}t)\,,\\
g_{1}&=& A(x)\frac{1}{8x^{4}}(-3+3x+4xt+5x^{2}t-6x^{3}t-8x^{3}t^{2})\,,\\
g_{2}&=& A(x)\frac{1}{16x^{4}}(3-3x-4xt+2x^{2}t)\,.
\end{eqnarray}
where $A(x)$ is given by
\begin{equation}
A(x) = A_{0}{\cal{Q}}^{2}\bigg(\frac{\mu_{7}^{2}(\alpha')^{4}}{\Lambda^{8}}\bigg)\bigg(\frac{\Lambda^{2}}{q^{2}}\bigg)^{\nu}
x^{\nu+5}(1-x)^{\nu-1}\,,
\end{equation}
and
$A_{0}=|c_{i}|^{2}|c_{X}|^{2}2^{4+2\nu}[\Gamma(2+\nu)]^{2}\pi^{5}$ is
a dimensionless normalization constant.

\section{DIS from mesons in the $\textmd{D4-D8-}\overline{\textmd{D8}}$-brane model}

In this section we extend the ideas of the previous one to the model
presented in \cite{Sakai:2003wu}. This model is based on a D-brane
construction in type IIA string theory consisting of a large number
of D4-branes and $N_f$ $\textmd{D8-}\overline{\textmd{D8}}$-brane
pairs in the probe approximation (though we will focus on the
Abelian flavour symmetry only). Direction 4 of the D4-branes is
compactified on an $S^{1}$, where antiperiodic boundary conditions
are imposed for fermions in the gauge theory, thus breaking all the
supersymmetries. This model geometrically realizes the
$U(N_{f})_{L}\times U(N_{f})_{R}$ chiral symmetry. The radial
coordinate $U$, which is transversal to the D4-branes has a lower
bound $U_{KK}$. The radius of $S^1$ shrinks to zero as $U
\rightarrow U_{KK}$. The D8 and the $\overline{\textmd{D8}}$ branes
merge at $U = U_{0} \ge U_{KK}$. In the resulting D8-brane there is
only a $U(N_{f})$ factor left. This is understood as a holographic
representation of the spontaneous breaking of the
$U(N_{f})_{L}\times U(N_{f})_{R}$ chiral symmetry down to
$U(N_{f})_V$.

\subsection*{The $\textmd{D4-D8-}\overline{\textmd{D8}}$-brane model}

We begin with the metric for the described system, details are given
in \cite{Sakai:2004cn},
\begin{eqnarray}\label{ss2}
&& ds^{2}=
\bigg(\frac{U}{R}\bigg)^{3/2}(\eta_{\mu\nu}dy^{\mu}dy^{\nu}+f(U)d\tau^{2})+
\bigg(\frac{R}{U}\bigg)^{3/2}\bigg(\frac {dU^{2}}{f(U)}+U^{2}
d\Omega_{4}^{2}\bigg)\:, \\
&& e^{\phi}=g_{s}\bigg(\frac{U}{R}\bigg)^{3/4}\:\:,\:\:F_{4}=dC_{3}=
\frac{2\pi
N_{c}}{V_{4}}\epsilon_{4}\:\:,\:\:f(U)=1-\frac{U^{3}_{KK}}{U^{3}}\:\:,
\end{eqnarray}
where $y^{\mu}$ ($\mu$= 0,1,2,3) and $\tau$ are the D4-brane
directions. The line element on the four-sphere is
$d\Omega_{4}^{2}$, the volume form is $\epsilon_{4}$ while the
$S^{4}$ volume is $V_{4}=8\pi^{2}/3$. Parameters $R$ and $U_{KK}$
are constants. $R$ is related to the string coupling $g_{s}$ and the
fundamental length $l_{s}$ by $R^{3}=\pi g_{s} N l_{s}^{3}$.

The $U$-coordinate has dimensions of length and is the radial
coordinate in the $56789$ directions which are transverse to the
D4-branes, $U\geq U_{KK}$.

The induced metric on a
D8-brane embedded in the above background is
\begin{equation}\label{ss9}
ds^{2}=
\bigg(\frac{U}{R}\bigg)^{3/2}\eta_{\mu\nu}dy^{\mu}dy^{\nu}+\bigg[\bigg(\frac{U}{R}\bigg)^{3/2}
f(U)+\bigg(\frac{R}{U}\bigg)^{3/2}\frac{U'^{2}}{f(U)}\bigg]d\tau^{2}+\bigg(\frac{R}{U}\bigg)^{3/2}U^{2}
d\Omega_{4}^{2} \,,
\end{equation}
with $U'=\frac{d}{d\tau}U$. The D8-brane action is proportional to
\begin{eqnarray}\label{ss10}
S_{D8} &\propto & \int d^{4}y d\tau\, \epsilon_{4}
e^{-\phi}\sqrt{|\textmd{det}(g_{D8})|} \propto \int d^{4}y d\tau
\,U^{4}\sqrt{f(U)+\bigg(\frac{R}{U}\bigg)^{3}\frac{U'^{2}}{f(U)}}\,.
\end{eqnarray}
Since the integrand does not explicitly depend on $\tau$ it is
possible to express $\tau$ as a function of $U$ and thus the metric
(\ref{ss9}) becomes
\begin{equation}\label{ss13}
ds^{2}=
\bigg(\frac{U}{R}\bigg)^{3/2}\eta_{\mu\nu}dy^{\mu}dy^{\nu}+\bigg(\frac{R}{U}\bigg)^{3/2}
\frac{U^{8}}{(U^{8} f(U)-U_{0}^{8}
f(U_{0}))}dU^{2}+\bigg(\frac{R}{U}\bigg)^{3/2}U^{2}
d\Omega_{4}^{2}\,.
\end{equation}
Now, let us recall that we want to study DIS from scalar and
polarized vector mesons in the context proposed in
\cite{Polchinski:2002jw}. We shall obtain the DIS structure
functions for $x\sim1$ (with $x<1$). The fundamental characteristic
of the geometry (\ref{ss13}) is the gravitational redshift, {\it
i.e.}  the warp factor multiplying the flat four-dimensional metric.
The momentum $p_{\mu}=-i\partial_{\mu}$ is identified with the one
in the gauge theory. As seen by an inertial observer the momentum on
the D-brane is
\begin{equation}\label{ss14}
\tilde{p}_{\mu}=\frac{1}{\sqrt{|g_{tt}|}}p_{\mu}\,.
\end{equation}
The characteristic energy scale on the D-brane is $R^{-1}$, thus the
four-dimensional energy scale is
\begin{equation}\label{ss15}
E^{(4)}\sim \frac{U^{3/4}}{R^{7/4}}\,.
\end{equation}
The dynamics of interest for DIS corresponds to the limit $q \gg
\Lambda$, where $\Lambda$ is the confinement energy scale of the
gauge theory. Thus, we shall consider the interaction in the
ultraviolet limit, being the interaction region given by
\begin{equation}\label{ss16}
U_{int}\sim q^{2} R^{3}\gg U_{0}= \Lambda^{2} R^{3} \equiv U_{KK}\,.
\end{equation}
In this region the metric (\ref{ss13}) can be approximated as
\begin{equation}\label{ss17}
ds^{2}=
\bigg(\frac{U}{R}\bigg)^{3/2}\eta_{\mu\nu}dy^{\mu}dy^{\nu}+\bigg(\frac{R}{U}\bigg)^{3/2}
dU^{2}+R^{3/2}U^{1/2} d\Omega_{4}^{2}\,.
\end{equation}
We shall study the matrix elements of two electromagnetic currents
in the hadron in order to extract the structure functions using the
optical theorem. Also, it is important to recall that as in the case
of D3-D7 model, the current inserted at the boundary induces the
excitation of a non-normalizable mode propagating in the bulk. The
boundary conditions for $A_m$ are given by
\begin{eqnarray}\label{ss57}
\lim_{U\rightarrow\infty}A_{\mu}(y, U)&=&n_{\mu}e^{iq \cdot y}=A_{\mu}(y)|_{4d}\,,\\
\lim_{U\rightarrow\infty}A_{U}(y, U)&=&0\,.
\end{eqnarray}
The gauge field satisfies the Maxwell equation $D_{m}F^{mn}=0$, in
the space given by $0123U$-coordinates with the metric (\ref{ss17}).
It is convenient to use the following gauge:
\begin{eqnarray}\label{ss51}
R^{-3}U^{3/4}\partial_{U}(U^{9/4}A_{U})+iq^{\nu}A_{\nu}=0\,.
\end{eqnarray}
We propose the Ans\"atze
\begin{eqnarray}\label{ss52}
A_{\mu}& \sim & n_\mu e^{iq \cdot y} \phi(U)\,,\\
A_{U} & \sim & e^{iq \cdot y} \varphi (U)\,,
\end{eqnarray}
then, the equations of motion become
\begin{eqnarray}\label{ss53}
U^{3/4}\partial_{U}(U^{9/4}\partial_{U}A_{\mu})-R^{3}q^{2}A_{\mu}=0\,,
\end{eqnarray}
\begin{eqnarray}\label{ss54}
R^{3}q^{2}A_{U}-\frac{3}{4}U^{-1/4}\partial_{U}(U^{9/4}A_{U})-U^{3/4}\partial_{U}^{2}(U^{9/4}A_{U})=0\,.
\end{eqnarray}
We obtain the solutions
\begin{eqnarray}\label{ss55}
A_{\mu}=\frac{2}{\Gamma(5/4)}n_{\mu}\:e^{iq \cdot
y}q^{5/4}\bigg(\frac{R^{3}}{U}\bigg)^{5/8}K_{5/4}\bigg(
\frac{2qR^{3/2}}{U^{1/2}}\bigg)\,,
\end{eqnarray}
\begin{eqnarray}\label{ss56}
A_{U}=-\frac{2iq \cdot n}{\Gamma(5/4)}\frac{q^{1/4}}{R^{3}}\:e^{iq
\cdot y}
\bigg(\frac{R^{3}}{U}\bigg)^{17/8}K_{1/4}\bigg(\frac{2qR^{3/2}}{U^{1/2}}\bigg)\,,
\end{eqnarray}
where $K_{5/4}$ and $K_{1/4}$ are modified Bessel functions, and $q
\equiv \sqrt{q^{2}}$.  The gauge fields satisfy the required
boundary conditions vanishing rapidly for $U<q^{2}R^{3}$, due to the
exponential behaviour of the Bessel functions. This implies that the
perturbation will be more suppressed in the interior of the space
defined by the $0123U$-coordinates as $q$ increases.

\subsection*{DIS from scalar mesons}

In this section we show how to obtain the scalar mesons in the
present model. Then, we obtain the interaction Lagrangian, and
finally the structure functions for DIS from scalar mesons, using
the method developed in the previous section. We begin with the
metric induced on the D8-brane \cite{Sakai:2004cn}
\begin{eqnarray}\label{ss58}
ds^{2}=
\bigg(\frac{U}{R}\bigg)^{3/2}\eta_{\mu\nu}dy^{\mu}dy^{\nu}+\bigg(\frac{R}{U}\bigg)^{3/2}
d\vec{Z}\cdot d\vec{Z}\,,
\end{eqnarray}
where $U=|\vec{Z}|$ and $Z^{i}$ ($i=1,\cdots,5$) parameterize the
56789-space. Scalar mesons correspond to fluctuations of the
D8-brane along the orthogonal directions. The dynamics of the
D8-brane is described by the following action
\begin{eqnarray}\label{ss59}
S_{D8}&=& -\mu_{8}\int
d^{9}\xi\sqrt{-\textmd{det}(P[G]_{ab}+2\pi\alpha'F_{ab})} +
\frac{(2\pi\alpha')^{2}}{2}\mu_{8}\int P[C^{(5)}]\wedge F\wedge F\,.
\end{eqnarray}
The D8-brane tension is
$\mu_{8}=[(2\pi)^{8}g_{s}\alpha'^{9/2}]^{-1}$ and $P$ denotes the
pullback of the background fields on the world-volume of the
D8-brane. The EOM for scalar mesons arises from the transversal
fluctuation
\begin{equation}\label{ss60}
x^{4}=L+2\pi \alpha' \Phi\,,
\end{equation}
being $\Phi$ the scalar fluctuation. From Eq.(\ref{ss59}) one
obtains the Lagrangian
\begin{equation}\label{ss61}
{\mathcal{L}}=-\mu_{8}\sqrt{|\textmd{det}(g_{ab})|}\sqrt{\textmd{det}(\delta^{ab}+R^{3/2}
U^{-3/2}g^{ac}(2\pi\alpha')^{2}\partial_{c}\Phi \partial^{b}
\Phi)}\, ,
\end{equation}
where all the indices denote directions on the D8-brane. Expanding
the Lagrangian up to quadratic order we obtain
\begin{eqnarray}\label{ss62}
{\mathcal{L}}=-\mu_{8}\sqrt{|\textmd{det}(g_{ab})|}\bigg[1+2\bigg(\frac{R}{U}\bigg)^{3/2}g^{ab}(\pi\alpha')^{2}
\partial_{a}\Phi \partial_{b} \Phi\bigg]\,.
\end{eqnarray}
Now, let us take spherical coordinates along the directions
($Z^{1}$, $\cdots$, $Z^{5}$), then use the metric (\ref{ss17}), plug
it into the quadratic Lagrangian, and obtain the following EOM
\begin{equation}\label{ss63}
\partial_{a} \bigg(\sqrt{\textmd{det}\tilde{g}}\,U^{7/4} g^{ab}\partial_{b}\Phi \bigg)=0\,,
\end{equation}
where $\tilde{g}_{ij}$ is the $S^{4}$-metric, which together with
the $U$ piece include directions ($Z^{1}$, $\cdots$, $Z^{5}$).
Expanding the EOM around the interaction region one obtains
\begin{eqnarray}\label{ss64}
R^{3}\partial^{\mu} \partial_{\mu}
\Phi+U^{-1/4}\partial_{U}(U^{13/4}\partial_{U}\Phi)+U\nabla^{i}\nabla_{i}\Phi=0\,,
\end{eqnarray}
where $\nabla_{i}$ is the covariant derivative on $S^{4}$. Indices
$i$, $j$, $k$, $\cdots$, denote coordinates on $S^{4}$ , $\mu$,
$\nu$, $\cdots$, denote directions parallel to the D4-branes and
$m$, $n$, $\cdots$, correspond to $0, 1, 2, 3, U$ coordinates. The
proposed Ansatz for the solution is
\begin{equation}\label{ss65}
\Phi=\phi(U)\:e^{iP \cdot y}\:Y^{\ell}(S^{4})\,,
\end{equation}
where $Y^{\ell}(S^{4})$ are the spherical harmonics on $S^{4}$,
\begin{equation}\label{ss65b}
\nabla^{i}\nabla_{i}Y^{\ell}(S^{4})=-\ell(\ell+3)Y^{\ell}(S^{4})\,.
\end{equation}

Now, the EOM for $\phi$ becomes
\begin{eqnarray}\label{ss70}
-P^{2}R^{3}\phi+U^{-1/4}\partial_{U}(U^{13/4}\partial_{U}\phi)-
m^{2}_{\ell} U \phi=0\,,
\end{eqnarray}
with $m^{2}_{\ell}=\ell(\ell+3)$. For the initial hadronic state we
can consider the leading behaviour of the solution for $U \sim
U_{int}$ where $|R^{3} P_{IN/OUT}^{2}| \ll U_{int}$. However, the
intermediate state $X$ must be represented by the full solution with
$P \equiv P_{X}$. Thus, we obtain
\begin{equation}\label{ss71}
\Phi_{IN/OUT}=c_{i}(\Lambda U)^{-9/8-\nu/2} e^{iP \cdot y}
Y(\Omega)\,,
\end{equation}
\begin{equation}\label{ss72}
\Phi_{X}=c_{X} \Lambda^{-13/8} s^{1/4} U^{-9/8}
J_{\nu}\bigg(\frac{2s^{1/2}R^{3/2}}{U^{1/2}}\bigg) e^{iP_{X} \cdot
y} Y(\Omega)\,,
\end{equation}
where $J_{\nu}$ is the Bessel function of first kind,
$\nu=(1/4)\sqrt{81+64 \ell(\ell+3)}$, being $c_{X}$ and $c_{i}$
dimensionless constants. If we analyze the parity, we can classify
the mesons as scalar and pseudoscalar mesons for even and odd values
of $\ell$, respectively since
$[Y^{\ell}(\Omega)]_{\textmd{P}}=(-1)^{\ell}Y^{\ell}(\Omega)$.

As in the  D3-D7 case, the fields (\ref{ss65}) are charged in the
space $0123U$. All the fields with $\ell > 0$ have charge
${\cal{Q}}$, since they transform with a phase under the $U(1)$
symmetry group induced by transformations on $S^{4}$ in the
direction of the Killing vector. Thus, the quadratic Lagrangian from
which we can derive the EOM (\ref{ss64}) is
\begin{equation}\label{ss67}
{\mathcal{L}}=-\mu_{8}(\pi \alpha')^{2}\sqrt{|\textmd{det}
g|}R^{3/2}U^{-3/2}g^{ab}(\partial_{a}\Phi
\partial_{b}\Phi^{*}+\partial_{b}\Phi \partial_{a}\Phi^{*})\,.
\end{equation}

The $U(1)$ current inserted at the boundary of the space spanned by
the $0123U$-coordinates, produces a perturbation on the boundary
conditions and excites a non-normalizable mode $A_{m}$, which
corresponds to the bulk extension of $A_{\mu}|_{4d}$. This field
couples to the $U(1)$ Noether's current in the bulk. The alluded
perturbation propagates in the bulk inducing a metric fluctuation
which is the product of $A_{m}(y, U)$ and a Killing vector
$\upsilon_{i}$ corresponding to the $U(1) \subset SO(5)$ isometry of
$S^{4}$. Therefore, we consider the fluctuation (\ref{pol5}),
$\delta g^{mi}=A^{m}(y, U)\upsilon^{i}(\Omega)$. Using the
eigenvalue equation
$\upsilon^{j}\partial_{j}Y(\Omega)=i{\cal{Q}}Y(\Omega)$, we obtain
the interaction Lagrangian
\begin{equation}\label{ss68}
{\mathcal{L}}_{int}=i{\cal{Q}}\mu_{8}(\pi
\alpha')^{2}\sqrt{|\textmd{det} g|}R^{3/2}U^{-3/2}A_{m}(\Phi
\partial^{m}\Phi^{*}_{X}-\Phi^{*}_{X}\partial^{m}\Phi )\,.
\end{equation}
Now, let us compute explicitly ${\cal{L}}_{int}$ from the coupling
of the gauge field $A_{m}$ to the Noether's current corresponding to
the internal symmetry $U(1)$ of the Lagrangian (\ref{ss67}). Under
an infinitesimal transformation with parameter
$\vartheta/2=\textmd{constant}$ we have
\begin{equation}\label{ss69}
j^{m}=i\mu_{8}(\pi \alpha')^{2}R^{3/2}U^{-3/2}(\Phi
\partial^{m}\Phi^{*}_{X}-\Phi^{*}_{X}\partial^{m}\Phi )\,.
\end{equation}
Thus, if we define ${\cal{L}}_{int}={\cal{Q}}\sqrt{|\textmd{det}
g|}A_{m}j^{m}$, we obtain the same ${\cal{L}}_{int}$ found in
Eq.(\ref{ss68}) from the metric fluctuation. Thus ${\cal{L}}_{int}$
is indeed given by the coupling of a gauge field to a conserved
current.

In order to obtain the coupling $A_{m}j^{m}$, we use the expressions
from (\ref{ss55}) and (\ref{ss56}),
\begin{eqnarray}
A_{\mu}&=& \frac{2}{\Gamma(5/4)}n_{\mu}\:e^{iq \cdot
y}q^{5/4}\bigg(\frac{R^{3}}{U}\bigg)^{5/8}K_{5/4}\bigg(
\frac{2qR^{3/2}}{U^{1/2}}\bigg)\,, \\
A_{U}&=& -\frac{i}{q^{2}}\eta^{\mu\nu}q_{\mu}\partial_{U}A_{\nu}\,,
\end{eqnarray}
and the current conservation
\begin{equation}
\partial_{\nu}j^{\nu}+U^{-13/4}\partial_{U}(U^{13/4}j^{U})=0\,,
\end{equation}
obtaining
\begin{equation}\label{ss73}
A_{m}j^{m}=A_{\mu}\bigg(j^{\mu}-i
\frac{q^{\mu}}{q^{2}}\partial_{\gamma}j^{\gamma}\bigg)-i\frac
{q^{\mu}}{q^{2}}U^{-13/4}\partial_{U}(U^{13/4}A_{\mu}j^{U})\,.
\end{equation}
Now, we obtain the action
\begin{eqnarray}\label{ss74}
S_{int}&=& \int_{U_{0}}^{\infty} d^{9}x{\cal{Q}}\sqrt{|\textmd{det} g|}A_{m}j^{m} \nonumber\\
&=& \int_{U_{0}}^{\infty} d^{9}x 2\mu_{8}
(\pi\alpha')^{2}{\cal{Q}}\sqrt{|\textmd{det}
\tilde{g}|}R^{15/4}U^{1/4}\bigg(P^{\mu}+\frac{q^{\mu}}{2x}\bigg)A_{\mu}
\Phi\Phi_{X}^{*}\nonumber\\ & & -i \frac
{q^{\mu}}{q^{2}R^{3/4}}\int d^{4}y
\:d\Omega{\cal{Q}}\sqrt{|\textmd{det}
\tilde{g}|}(A_{\mu}U^{13/4}j^{U})|^{\infty}_{\Lambda^{2}
R^{3}}\,.\:\:\:\:\:
\end{eqnarray}
Using $z=R^{3/2}/U^{1/2}$ and $1/\Lambda\rightarrow\infty$, since
$A_{\mu}$ goes to zero very fast for $U<U_{int}$, the result is
\begin{equation}\label{ss75}
\langle P+q, X|J^{\mu}(0)|P, {\cal {Q}}\rangle \,=\kappa_3
I_{3}\bigg(P^{\mu}+\frac{q^{\mu}}{2x}\bigg)\,,
\end{equation}
where $c_{i}$ has been redefined,
$\kappa_{3}=8\mu_{8}{\cal{Q}}(\pi\alpha')^{2}\:c_{i}c_{X}^{*}[\Gamma(5/4)]^{-1}q^{7/4}\Lambda^{\nu-7/2}x^{-1/4}
(1-x)^{1/4}$ and the integral in $U$ gives $I_{3}$, which using the
$s\simeq q^{2}(1/x-1)$ approximation becomes
\begin{equation}
I_{3}=\frac{1}{4}\Gamma\bigg(\frac{9}{4}+\nu\bigg)q^{-\nu-13/4}x^{9/4+\nu/2}(1-x)^{\nu/2}\,.
\end{equation}
In order to compute $\textmd{Im}\:T^{\mu\beta}$ from
Eq.(\ref{pol3}), we need to multiply Eq.(\ref{ss75}) by its complex
conjugate and proceed in an analogous way as before with the D3-D7
system. The distance between zeros of the Bessel functions in
(\ref{ss72}) is $M_{n}=n\pi\Lambda$. The density of states is
obtained in the same way as in the previous section. The final
result of the tensor is
\begin{eqnarray}\label{ss77}
\textmd{Im} \:T^{\mu\beta}&=&  8\pi^{5}{\cal{Q}}^{2}[\Gamma(9/4+\nu)]^{2}
[\Gamma(5/4)]^{-2}\mu_{8}^{2}(\alpha')^{4}|c_{i}|^{2}|c_{X}|^{2} \nonumber\\ & & \times \Lambda^{2\nu-8}
q^{-4-2\nu}(1-x)^{\nu}x^{\nu+9/2}\bigg(P^{\mu}+\frac{q^{\mu}}{2x}\bigg)
\bigg(P^{\beta}+\frac{q^{\beta}}{2x}\bigg)\,.\,\,\,\,\,\,\,\,\,\,\,\,\,\,\,\,\,\,\,\,\,\,
\end{eqnarray}
We obtain
\begin{equation}\label{ss78}
F_{1}=0, \,\,\,\,\,\,\,\,\,\:\:\:\:F_{2}=E'_{0}{\cal{Q}}^{2}
\bigg(\frac{\mu_{8}^{2}\alpha'^{4}}{\Lambda^{10}}\bigg)
\bigg(\frac{\Lambda^{2}}{q^{2}}\bigg)^{\nu+1}x^{\nu+7/2}(1-x)^{\nu}\,,
\end{equation}
where
$E'_{0}=4\pi^{5}|c_{i}|^{2}|c_{X}|^{2}[\Gamma(9/4+\nu)]^{2}[\Gamma(5/4)]^{-2}$
is a normalization dimensionless constant.

\subsection*{DIS from vector mesons}

In this section we obtain vector mesons in the Sakai-Sugimoto model,
compute the ${\cal{L}}_{int}$ and derive the structure functions of
DIS from polarized spin-one mesons in the context of the type IIA
string dual description. In fact, we have obtained a full tower of
vector mesons from the Dirac-Born-Infeld action of a D8-brane
($N_{f}=1$) considering fluctuations along the brane, which has an
additional quantum number $\ell$ with respect to the mesons
calculated in \cite{Sakai:2004cn}. We are specially interested in
mesons with $\ell>0$ as we need to study charged fields in order to
be able to derive the interaction Lagrangian. So, for $\ell>0$, our
mesons will transform with a phase under the $U(1)$ symmetry induced
by transformations on the $S^{4}$ in the direction of a Killing
vector. This can be satisfied by requiring the mesons not only to be
dependent on the coordinates $0,1,2,3,U$, but they should also
depend on the angular coordinates of the $S^{4}$. As a result, our
mesons are charged and massive from the point of view of the
five-dimensional space obtained after dimensional reduction on the
$S^{4}$.

Besides, due to the fact that the relevant interaction region for
the DIS corresponds to large values of the four-dimensional energy,
we approximate the metric for that specific region. Consequently, we
have calculated full analytic expressions for these mesons depending
on the nine coordinates of the D8-brane.

From the action (\ref{ss59}) we derive the EOM, then we propose a
quadratic Lagrangian from which we obtain exactly the same EOM, thus
\begin{eqnarray}\label{ss79}
{\cal{L}}_{D8}&=&
-\mu_{8}\sqrt{-\textmd{det}(g_{ab}+2\pi\alpha'F_{ab})}+
\frac{(2\pi\alpha')^{2}}{2}\mu_{8}P[C^{(5)}]\wedge F\wedge F \,,
\,\,\,\,\,\,\,\,\,\,\,\,\,\,\,\,\,\,\,\,\,\,\,\\
&=& {\cal{L}}_{DBI}+{\cal{L}}_{CS}\,.\nonumber
\end{eqnarray}
The Ansatz for the solution of vector mesons is
\begin{eqnarray}\label{ss81}
B_{\mu}=\zeta_{\mu}\:\phi(U)\:e^{iP \cdot
y}\:Y^{\ell}(S^{4})\,,\:\:\:\:\:P \cdot
\zeta=0\,,\:\:\:\:\:B_{U}=0\,,\:\:\:\:\:B_{i}=0\,,
\end{eqnarray}
where we have expanded it in spherical harmonics of $S^{4}$ (there
is a field $B_{\mu}$ for each $\ell$, as for the D3-D7 brane
system). We have to determine $\phi(U)$. The polarization vector is
$\zeta_{\mu}$ and $P \cdot \zeta=0$ comes from
$\partial^{\mu}B_{\mu}=0$. Taking into account the Ansatz above the
EOM reads
\begin{eqnarray}\label{ss80}
\partial_{a}(\sqrt{|\textmd{det} g|}F^{ab})=0\,,
\end{eqnarray}
where $a$, $b$, $c$, $d$, $\cdots$ run over all the coordinates of
the D8-brane.

Plugging the Ansatz in Eq.(\ref{ss80}), we find that solutions with
$b=k$ and $b=U$ identically satisfy the EOM from
$\partial^{\mu}B_{\mu}=0$. If $b=\mu$ we find
\begin{eqnarray}\label{ss82}
R^{3}U^{-3}\eta^{\alpha \beta} \eta^{\mu \nu}(P_{\alpha}P_{\nu}B_{\beta}
-P_{\alpha}P_{\beta}B_{\nu})+U^{-13/4}\eta^{\mu \nu} \partial_{U}(U^{13/4}
\partial_{U}B_{\nu})-U^{-2}\eta^{\mu \nu} m_{\ell}^{2}B_{\nu}&=&0\,,\:\:\:\:\:\:\:\:\:\:\:\:\:
\end{eqnarray}
which leads to an equation equal to Eq.(\ref{ss70}) for $\phi(U)$
while $m^{2}_{\ell}=\ell(\ell+3)$ comes from
\begin{eqnarray}\label{ss83}
\nabla^{2}_{S^{4}}Y^{\ell}(S^{4})=-\ell(\ell+3)Y^{\ell}(S^{4})\, .
\:\:\:
\end{eqnarray}
The solution is
\begin{eqnarray}\label{ss85}
\phi(U)=c\,
U^{-9/8}J_{\nu}\bigg(\frac{2s^{1/2}R^{3/2}}{U^{1/2}}\bigg)\,,
\end{eqnarray}
where $J_{\nu}$ is the Bessel function of first kind, $c$ being a
dimensionless constant, while $\nu=(1/4)\sqrt{81+64\ell(\ell+3)}$.
Again, as in the scalar case, for initial and final hadronic states
we can use the leading behaviour for $|P^{2}| \ll q^{2}$, then
\begin{eqnarray}\label{ss86}
B_{\mu \,IN/OUT}=\zeta_{\mu}c_{i}\Lambda^{-1}(\Lambda
U)^{-\nu/2-9/8}e^{iP \cdot y}\:Y^{\ell}(S^{4})\,,
\end{eqnarray}
being $c_{i}$ a dimensionless constant. The intermediate state is
given by the full solution
\begin{eqnarray}\label{ss87}
B_{X \mu}=\zeta_{X \mu}c_{X}\Lambda^{-13/8}s^{-1/4}
U^{-9/8}J_{\nu}\bigg(\frac{2s^{1/2}R^{3/2}}{U^{1/2}}\bigg)e^{iP_{X}.y}\:Y^{\ell}(S^{4})\,,
\end{eqnarray}
where $c_{X}$ is a dimensionless constant. The radius of the Bessel
function $2s^{1/2}R^{3/2}$ is much larger than $U_{0}$, being this
the reason to consider the full solution for intermediate states. To
study the parity transformations, using the relations
$[Y^{\ell}(\Omega)]_{\textmd{P}}=(-1)^{\ell}Y^{\ell}(\Omega)$ and
$[\zeta_{\mu}]_{\textmd{P}}=(\zeta^{0}, -\vec{\zeta})$ we can
classify the mesons as vector mesons for even values of $\ell$ and
axial vector mesons for odd values of $\ell$.

From the expansion of $B_{\mu}$ in spherical harmonics of $S^{4}$,
it can be seen that the gauge fields on the D8-brane correspond to
charged fields in the space defined by $0123U$ coordinates in an
analogous way as for the D3-D7-brane model. The quadratic Lagrangian
which gives the EOM (\ref{ss82}) for the solution (\ref{ss81}) of
vector mesons is
\begin{eqnarray}\label{ss88}
{\mathcal{L}}=-2\mu_{8}(\pi \alpha')^{2}\sqrt{|\textmd{det}
g|}F^{ab}F_{ab}^{*}\,.
\end{eqnarray}
Considering a fluctuation of the metric of the form $\delta
g_{mj}=A_{m}(y,U)\upsilon_{j}(\Omega)$ with
$\upsilon^{j}\partial_{j}Y(\Omega)=i{\cal{Q}}Y(\Omega)$ the
interaction Lagrangian is:
\begin{equation}\label{ss89}
{\mathcal{L}}=2i{\cal{Q}}\mu_{8}(\pi \alpha')^{2}\sqrt{|\textmd{det}
g|}A_{m}[B_{X\mu}^{*}F^{\mu m}-B_{\mu}(F_{X}^{\mu m})^{*}]\, ,
\end{equation}
which can alternatively be obtained from the coupling of the gauge
field to the Noether's current corresponding to the $U(1)$ symmetry
as before, being ${\cal{L}}_{int}={\cal{Q}}\sqrt{|\det
g|}A_{m}j^{m}$ while the current is given by
\begin{equation}\label{ss90}
j^{m}=2i\mu_{8}(\pi \alpha')^{2}[B_{X\mu}^{*}F^{\mu
m}-B_{\mu}(F^{\mu m}_{X})^{*}]\,.
\end{equation}
Then, the coupling of $A_{m}$ to a conserved current $j^{m}$ leads
to ${\cal{L}}_{int}$.

In order to obtain the gauge field coupled to the conserved current,
we use a similar procedure to that in the previous section, thus
\begin{eqnarray}\label{ss91}
S_{int}&=& \int_{U_{0}}^{\infty} d^{9}x{\cal{Q}}\sqrt{|\textmd{det} g|}A_{m}j^{m} \nonumber\\
&=& \int_{U_{0}}^{\infty} d^{9}x \,2\mu_{8}
(\pi\alpha')^{2}{\cal{Q}}\sqrt{|\textmd{det} g|} R^{3} U^{-3}
A_{\mu}N^{\mu} \tilde{B}\tilde{B}_{X}^{*}\nonumber\\ & & -i \frac
{q^{\mu}R^{3/4}}{q^{2}}\int d^{4}y d\Omega\sqrt{|\textmd{det}
\tilde{g}|}(A_{\mu}U^{13/4}j^{U})|^{\infty}_{\Lambda^{2}
R^{3}}\,,\:\:\:\:\:
\end{eqnarray}
where $B_{\mu}=\zeta_{\mu}\tilde{B}$ and
\begin{equation}\label{ss92}
N^{\mu}=2(\zeta \cdot
\zeta_{X})\bigg(P^{\mu}+\frac{q^{\mu}}{2x}\bigg)+(\zeta_{X} \cdot
q)\zeta^{\mu}-(\zeta \cdot q)\zeta^{\mu}_{X}\,.
\end{equation}
Considering $z=R^{3/2}/U^{1/2}$ and $1/\Lambda\rightarrow\infty$
since $A_{\mu}$ goes to zero very fast for $U<U_{int}$, the second
term in Eq.(\ref{ss91}) vanishes when evaluated at the integration
limits. Thus by integration of the remaining term and comparing with
Eq.(\ref{kru21.1}), we obtain
\begin{equation}\label{ss93}
\langle P+q, X|J^{\mu}(0)|P, {\cal {Q}} \rangle \,= \kappa_{4} I_{4}
N^{\mu}\,,
\end{equation}
where $\kappa_{4}=8\mu_{8}{\cal{Q}}(\pi\alpha')^{2}c_{i}c_{X}^{*}
[\Gamma(5/4)]^{-1}\Lambda^{\nu-9/2} \:q^{3/4}x^{1/4}(1-x)^{-1/4}$
and the result of the integral in $U$ is $I_{4}$, that in the
approximation $s\simeq q^{2}(1/x-1)$ corresponds to
\begin{eqnarray}
I_{4} &=& 2^{-2}\Gamma(\nu+9/4)q^{-\nu-13/4}
x^{\nu/2+9/4}(1-x)^{\nu/2}\,,
\end{eqnarray}
In order to compute $\textmd{Im}\:T^{\mu\beta}$ given in
(\ref{pol3}) we have to multiply Eq.(\ref{ss93}) by its complex
conjugate, add the radial excitations, and finally sum over the
polarizations of the final hadronic states, obtaining
\begin{eqnarray}\label{ss94}
\textmd{Im} \:T^{\mu\beta}&=& \frac{\pi\kappa_{4}\kappa_{4}^{*}I_{4}^{2}}{\Lambda s^{1/2}}\sum_{\lambda}N^{\mu}N^{*\beta}\nonumber\\
&=&\frac{\pi\kappa_{4}\kappa_{4}^{*}I_{4}^{2}}{\Lambda
s^{1/2}}\sum_{\lambda}\bigg\{\bigg[2(\zeta \cdot
\zeta_{X})\bigg(P^{\mu}+\frac{q^{\mu}}{2x}\bigg)+(\zeta_{X} \cdot q)
\zeta^{\mu}-(\zeta \cdot q)\zeta^{\mu}_{X}\bigg]\nonumber\\& &
\times \bigg[ 2(\zeta^{*} \cdot
\zeta^{*}_{X})\bigg(P^{\beta}+\frac{q^{\beta}}{2x}\bigg)+(\zeta^{*}_{X}
\cdot q) \zeta^{*\beta}-(\zeta^{*} \cdot
q)\zeta^{*\beta}_{X}\bigg]\bigg\}\,.
\end{eqnarray}
Also, for the solution (\ref{ss81}) it holds $P \cdot\zeta=0$, which
implies that there are three polarizations $\lambda=1, 2, 3$. We use
the normalization for the polarizations given by $\zeta^{\mu}(P_{X},
\lambda) \cdot \zeta_{\mu}^{*}(P_{X},
\lambda')=-M^{2}_{X}\delta_{\lambda, \lambda'}$, and we obtain
\begin{eqnarray}\label{ss95}
\sum_{\lambda}\zeta_{X\mu}(P_{X},\lambda)\zeta_{X\beta}^{*}(P_{X},\lambda)&=& M^{2}_{X}
\eta_{\mu\beta}+P_{X\mu}P_{X\beta}\,,\nonumber\\
&=&
M^{2}_{X}\eta_{\mu\beta}+(P_{\mu}+q_{\mu})(P_{\beta}+q_{\beta})\,.\nonumber
\end{eqnarray}
The rest of the computation is similar to the one done in the
previous section for the D3-D7-brane system. We introduce the final
form for the structure functions in the Sakai-Sugimoto model
\cite{Sakai:2004cn} which we have obtained:
\begin{eqnarray}
F_{1}&=& E(x) \frac{1}{12x^{3}}(1-x-2xt-4x^{2}t+4x^{3}t+8x^{3}t^{2})\,,\\
F_{2}&=& E(x) \frac{1}{6x^{3}}(1-x+12xt-14x^{2}t-12x^{2}t^{2})\,,\\
b_{1}&=& E(x) \frac{1}{4x^{3}}(1-x-xt)\,,\\
b_{2}&=& E(x) \frac{1}{2x^{3}}(1-x-x^{2}t)\,,\\
b_{3}&=& E(x) \frac{1}{24x^{3}}(1-4x+8x^{2}t)\,,\\
b_{4}&=& E(x) \frac{1}{12x^{3}}(-1+4x-2x^{2}t)\,,\\
g_{1}&=& E(x) \frac{1}{8x^{4}}(-3+3x+4xt+5x^{2}t-6x^{3}t-8x^{3}t^{2})\,,\\
g_{2}&=& E(x) \frac{1}{16x^{4}}(3-3x-4xt+2x^{2}t)\,.
\end{eqnarray}
We can observe that they have the same form as the ones obtained
with the D3-D7-model, however the factor $E(x)$, which turns out to be
common to all the structure functions, depends on the model and
differs with respect to the previous model. $E(x)$ is given by
\begin{equation}
E(x) = E_{0}{\cal{Q}}^{2}\bigg(\frac{\mu_{8}^{2}(\alpha')^{4}}{\Lambda^{10}}\bigg)
\bigg(\frac{\Lambda^{2}}{q^{2}}\bigg)^{\nu}
x^{\nu+11/2}(1-x)^{\nu-1}\,,
\end{equation}
where
$E_{0}=\pi^{5}4|c_{i}|^{2}|c_{X}|^{2}[\Gamma(\nu+9/4)]^{2}[\Gamma(5/4)]^{-2}$
is a dimensionless normalization constant.

\section{Results and Discussion}

In this section we discuss our results extensively. We have
investigated deep inelastic scattering from scalar mesons and
polarized vector mesons obtained from two holographic dual models
with flavours in the fundamental representation of the gauge group.
These models are the $\textmd{D3-D7}$-brane model
\cite{Kruczenski:2003be} and the
$\textmd{D4-D8-}\overline{\textmd{D8}}$-brane model
\cite{Sakai:2004cn}, which are dual to the large $N$ limit of
$SU(N)$ ${\cal {N}}=2$ supersymmetric Yang Mills theory and of
$SU(N)$ non-supersymmetric Yang-Mills theory within the same
universality class as QCD\footnote{There are certain differences
with the large $N$ limit of QCD itself. For instance, the model in \cite{Sakai:2004cn} has a global
$SO(5)$ symmetry which QCD does not have.}, respectively. Although
we have only considered one flavour ($N_f=1$), the method described
above can be straightforwardly extended to the case of $N_f>1$.

As discussed above, DIS amplitudes are obtained from the Lorentz
contraction of a leptonic tensor with a hadronic tensor $W^{\mu\nu}$
which characterizes the structure of the hadron. Since strongly
coupled effects -which cannot be accounted for by using quantum
field theory perturbative methods- contribute to this tensor, it is
necessary to approach the problem in a completely different way. In
fact, this is achieved through the implementation of a holographic
dual description, which makes it possible to investigate the
strongly coupled regime of the field theory in terms of a string
theory dual model. With this in mind, we have derived the hadronic
tensor using two different holographic dual models, in the large $N$
limit and for large values of the 't Hooft coupling. In this regime,
as suggested by Polchinski and Strassler \cite{Polchinski:2002jw},
the currents are scattered by the entire hadron rather than by any
constituent parton. We shall comment more on this point.

Let us briefly summarize a few properties satisfied by the
holographic dual models we have considered. The interaction
Lagrangian used in both models is invariant under parity and time
reversal symmetries. Also we ought to mention that the existence of
a conserved current in the holographic dual model implies that
$q_\mu W^{\mu\nu} = q_\nu W^{\mu\nu} = 0$ in the four-dimensional
theory, where $q_\mu$ is the virtual photon four-momentum. In four
dimensions this property comes from the conservation of the current
coupled to the gauge field in the interaction Lagrangian. In order
to obtain the interaction Lagrangian in the holographic dual models
we have implemented a systematic method based upon the computation
of the Noether's current of the DBI Lagrangian expanded up to
quadratic order in the derivatives of the fields. By applying the
method based on the Noether's current, we have found a ${\cal
{L}}_{int}$ with similar properties for both models, {\it i.e.} the
$\textmd{D3-D7}$-brane model and the
$\textmd{D4-D8-}\overline{\textmd{D8}}$-brane model. This guarantees
the existence of a conserved current and, thus $q_\mu W^{\mu\nu} =
q_\nu W^{\mu\nu} = 0$ is automatically satisfied.

If we analyze the parity transformations in both models studied, we
can classify the mesons in towers of scalar, pseudoscalar, vector
and axial vector mesons. This classification depends on the values
of $\ell$, which characterizes the spherical harmonics,
$Y^{\ell}(\Omega)$, of the mesons wave function. It is interesting
to stress that within the $\textmd{D4-D8-}\overline{\textmd{D8}}$
model we have obtained analytic expressions for scalar as well as
for vector mesons which are massive in the five-dimensional
spacetime defined by the coordinates $(y^{0}, y^{1}, y^{2}, y^{3},
U)$. Those expressions hold for large values of the energy of the
dual four-dimensional quantum field theory, given by $E^{(4)}\sim
U^{3/4}/R^{7/4}$, which corresponds to $U \gg U_{0}$, where
$U_{0}=\Lambda^2 R^3$ is the minimum of the coordinate $U$. The
tower of vector fluctuations of the D8-brane that we have calculated
has an additional quantum number, $\ell$, compared to the mesons
considered in \cite{Sakai:2004cn}, that comes from the expansion in
spherical harmonics on $S^4$.

We have obtained the DIS structure functions for the whole scalar
and vector meson spectra using both holographic dual models, getting
exact analytical expressions for these functions in all the cases.
Particularly, for the vector mesons we have got the full tensor
structure decomposition of the hadronic tensor for polarized targets
of spin-1. In fact, we have computed all the structure functions:
$F_{1},\, F_{2},\, b_{1}, \,b_{2}, \,b_{3},\, b_{4}, \,g_{1}$ and
$g_{2}$, as functions of $x$ and $t$. This problem had not been
considered in the previous literature of vector mesons using the
gauge/string duality, so this is a totally new approach to the
problem which complements those based on the operator product
expansion of electromagnetic currents and the parton model approach
investigated by Hoodbhoy, Jaffe and Manohar \cite{Hoodbhoy:1988am}.
The present calculation gives nontrivial results and predictions for
the structure functions of spin-1 hadrons, particularly focussed on
vector mesons.

We find that $F_1=0$ for scalar mesons as expected. Recall that
$F_1$ is proportional to the Casimir of the scattered hadron under
the Lorentz group and therefore it vanishes for a scalar hadron as
it would do for a scalar parton, in agreement with our result above.
On the other hand, $F_2$ provides spin-independent information and
does not vanish. Besides, for the scalar mesons, the dependence on
momentum squared in the equation for $F_2$, which is $F_2 \sim
q^{-2(\tau_{p}-1)}$, corresponds to the twist $\tau_{p}=\nu+2$ in
our notation (note that $\nu$ is different in each model). For the
vector mesons, we obtain finite results for all the structure
functions, that we discuss below.

Below we display our results about the structure functions in
Figures (2-4), corresponding to the $\textmd{D3-D7}$-brane model
(left) and $\textmd{D4-D8-}\overline{\textmd{D8}}$-brane model
(right). In all figures we have considered only the lowest values of
the mass $m_\ell$, which depends on the model. In these figures only
the dependence upon the Bjorken parameter is shown. In particular,
the following functions are displayed\footnote{In Figures (2-4) we
show the structure functions obtained in the previous two sections
re-scaled as Eq.(\ref{d1}). Notice that we use the notation
$F_{1},\, F_{2}, \, b_{1}, \, b_{2}, \, b_{3}, \, b_{4}, \, g_{1},
\, g_{2}$ for the corresponding re-scaled structure functions. Also,
the functions are multiplied by different factors (powers of 10
explicitly written in Eq.(\ref{d1})) in each model in order to make
the comparison between the two models easier. }:
\begin{eqnarray}\label{d1}
& & F_{2}\,\times 10^{3}/\bigg[A'_{0}{\cal{Q}}^{2}
\bigg(\frac{\mu_{7}^{2}(\alpha')^{4}}{\Lambda^{8}}\bigg)
\bigg(\frac{\Lambda^{2}}{q^{2}}\bigg)^{\nu+1}\bigg]\,\,\,\,\,\,\,\,\,\,\,\,\,\,\,\,\,\,\,\,\,\,\,\,\,\,\,\,\,
\,\textmd{for D3-D7 scalar mesons}\,,\nonumber\\
& & F_{2}\,\times 10^{6}/\bigg[E'_{0}{\cal{Q}}^{2}
\bigg(\frac{\mu_{8}^{2}(\alpha')^{4}}{\Lambda^{10}}\bigg)
\bigg(\frac{\Lambda^{2}}{q^{2}}\bigg)^{\nu+1}\bigg]\,\,\,\,\,\,\,\,\,\,\,\,\,\,\,\,\,\,\,\,\,\,\,\,\,\,\,\,\,\,\textmd{for
D4-D8-}\overline{\textmd{D8}}
\, \textmd{scalar mesons} \,,\nonumber\\
& &(F_{i},\,b_{j},\,g_{i})\times
10^{3}/\bigg[A_{0}{\cal{Q}}^{2}\bigg(\frac{\mu_{7}^{2}(\alpha')^{4}}{\Lambda^{8}}\bigg)
\bigg(\frac{\Lambda^{2}}{q^{2}}\bigg)^{\nu}\bigg]\,\,\,\,\,\,\,\,\,\,\,\,\,\,\,\,\textmd{for D3-D7 vector mesons}\,,\nonumber\\
& &(F_{i},\,b_{j},\,g_{i})\times
10^{6}/\bigg[E_{0}{\cal{Q}}^{2}\bigg(\frac{\mu_{8}^{2}(\alpha')^{4}}{\Lambda^{10}}\bigg)
\bigg(\frac{\Lambda^{2}}{q^{2}}\bigg)^{\nu}\bigg]
\,\,\,\,\,\,\,\,\,\,\,\,\,\,\,\,\textmd{for D4-D8-}\overline{\textmd{D8}} \, \textmd{vector mesons}\,,\nonumber\\
\end{eqnarray}
where $i=1, 2$, $j=1, 2, 3, 4$ and $\nu=\ell+1$ for the D3-D7 model
and $\nu=(1/4)\sqrt{81+64\ell(\ell+3)}$ for the
D4-D8-$\overline{\textmd{D8}}$ model. In these figures we have
neglected terms proportional to $t$ with respect to the ones
proportional to $x^{n}$ with $n\geq 0$, since the computation has
been done within the regime where  $x \sim 1$ (although horizonal
axes include the full range $0 < x \leq 1$ we assume our
calculations to hold for $x \sim 1$) and $|t|= (-P^{2}/q^{2})\ll1$,
since $-P^{2}\ll q^{2}$ ($P$ is the ingoing meson four-momentum).

For the scalar mesons, from Figure 2 it can be seen that the Bjorken
parameter dependence of $F_{2}$ is similar in both models. This
similarity between the structure functions in both models is
extended to the vector mesons, as shown in Figures 3 and 4. The
reason is that each structure function in both models shares a
common factor which is model-independent and is determined by their
corresponding DBI actions. However, each of them has a
model-dependent factor, hence the differences between curves in both
models.

For the vector mesons, we have found the modified Callan-Gross
relation as previously obtained in \cite{Polchinski:2002jw}, namely
$2 F_1(x)=F_2(x)$, which holds for every value of the Bjorken
parameter in the present case. In reference \cite{Polchinski:2002jw}
this modified Callan-Gross relation was obtained for the case of
spin-$\frac{1}{2}$ hadrons, corresponding to supergravity modes of
the dilatino. In the present case we obtain the same relation for
the supergravity modes of gauge fields which are fluctuations along
the flat directions of the D7 and D8 brane world-volumes, depending
on the model we consider. As pointed out in \cite{Polchinski:2002jw}
the difference with respect to the usual Callan-Gross relation is an
$x$-factor missing on the left-hand side of the relation. In the
case of a scattered parton which carries a fraction $x$ of the total
four-momentum of the hadron, the operator product expansion leads to
$2 x F_1(x)=F_2(x)$. On the other hand, in the pure supergravity
description the entire hadron is expected to be scattered due to the
strong coupling limit in which the system is studied. We emphasize
that within the supergravity approximation our result $2
F_1(x)=F_2(x)$ is exact\footnote{For this relation as well as for
the ones that are considered below, terms proportional to $t$ have
been neglected.}. In addition, we also find the exact relation $2
b_1(x)=b_2(x)$, which is the pure supergravity version of $2 x
b_1(x)=b_2(x)$, also obtained from the operator product expansion
analysis of the time ordered product of two electromagnetic
currents. The last relation comes from the fact that $b_1$ and $b_2$
are obtained from the same operator tower as $F_1$ and $F_2$, thus
they obey the same scaling equations. Also notice that in the parton
model $F_1$ represents the probability of finding a quark in the
hadron with momentum fraction $x$, while $g_1$ is the difference
between the probabilities of finding a quark in the hadron with
momentum fraction $x$ with spin parallel and antiparallel to the
hadron \cite{Manohar:1992tz}. Thus, in principle, $g_1$ can be
positive or negative. Our results agree with these physical
interpretations.
\begin{figure}
\begin{center}
\epsfig{file=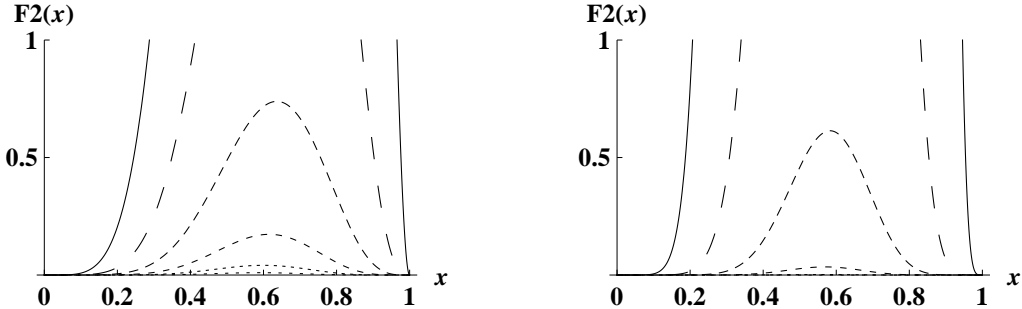, width=15cm}{\caption{\small
Structure function $F_2(x)$ for scalar mesons, scaled as explained
in the text, as a function of the Bjorken parameter. Left and right
figures correspond to the $\textmd{D3-D7}$-brane and the
$\textmd{D4-D8-}\overline{\textmd{D8}}$-brane models, respectively.
Different curves stand for different values of $\ell > 0$, we
display the lowest values of it. The curve sizes increase with
$\ell$. Notice that $t$ has been neglected.}}
\label{FigF2ScalarMesons}
\end{center}
\end{figure}
\begin{figure}
\begin{center}
\vspace{0.5cm}
\epsfig{file=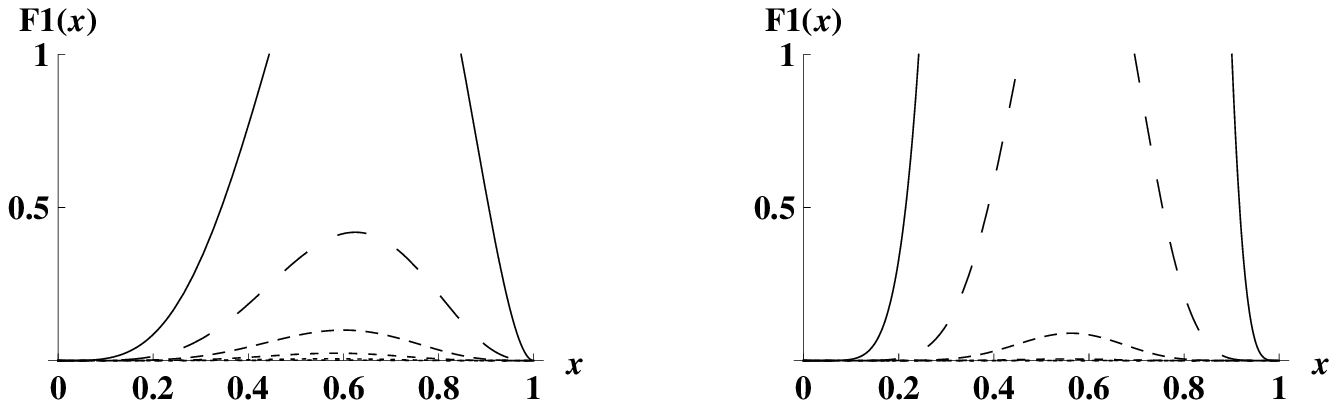, width=15cm}
\vspace{0.5cm}
\epsfig{file=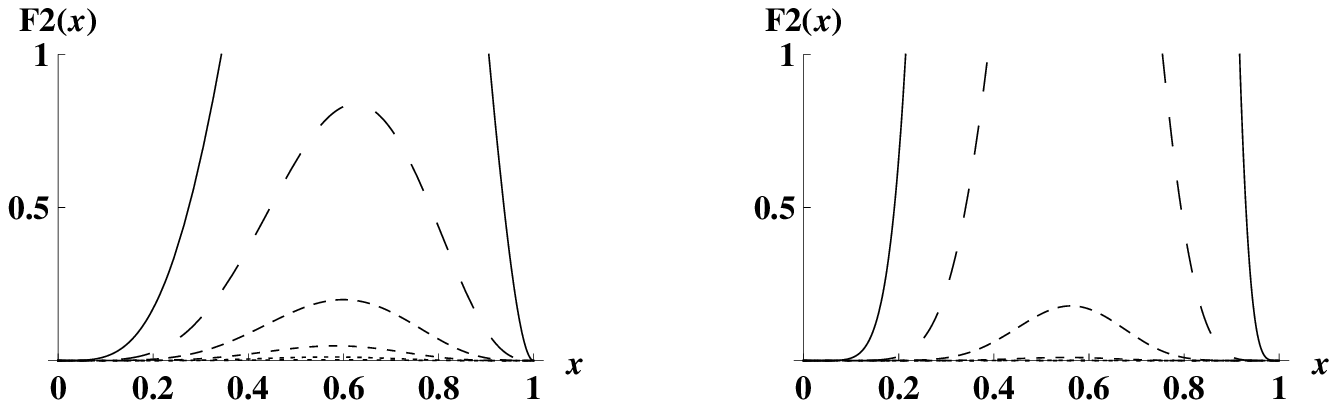, width=15cm}
\vspace{0.5cm}
\epsfig{file=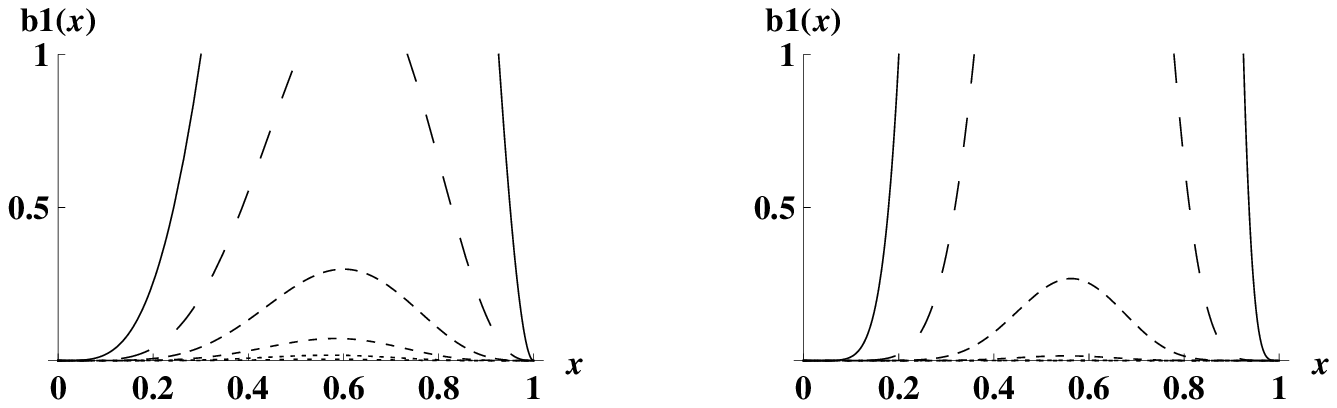, width=15cm}
\vspace{0.5cm}
\epsfig{file=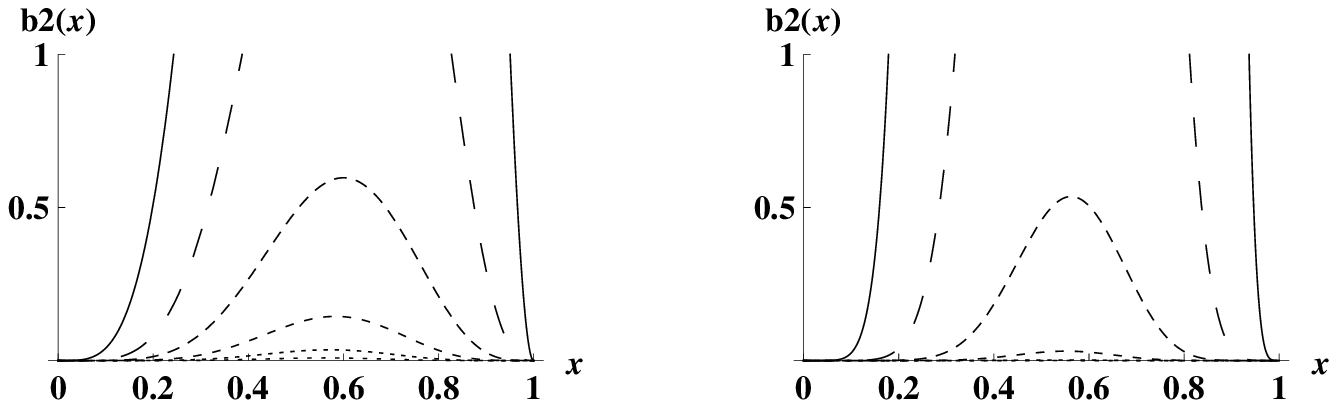, width=15cm} {\caption{\small
Structure functions $F_1(x)$, $F_2(x)$, $b_1(x)$ and $b_2(x)$ for
vector mesons as a function of the Bjorken parameter scaled as
explained in the text. Left and right figures correspond to the
$\textmd{D3-D7}$-brane and the
$\textmd{D4-D8-}\overline{\textmd{D8}}$-brane models, respectively.
Different curves stand for different values of $\ell > 0$, we
display the lowest values of it. The curve sizes increase with
$\ell$. Notice that $t$ has been neglected. }}
\label{Figb2vectormesons}
\end{center}
\end{figure}
\begin{figure}
\begin{center}
\vspace{0.5cm}
\epsfig{file=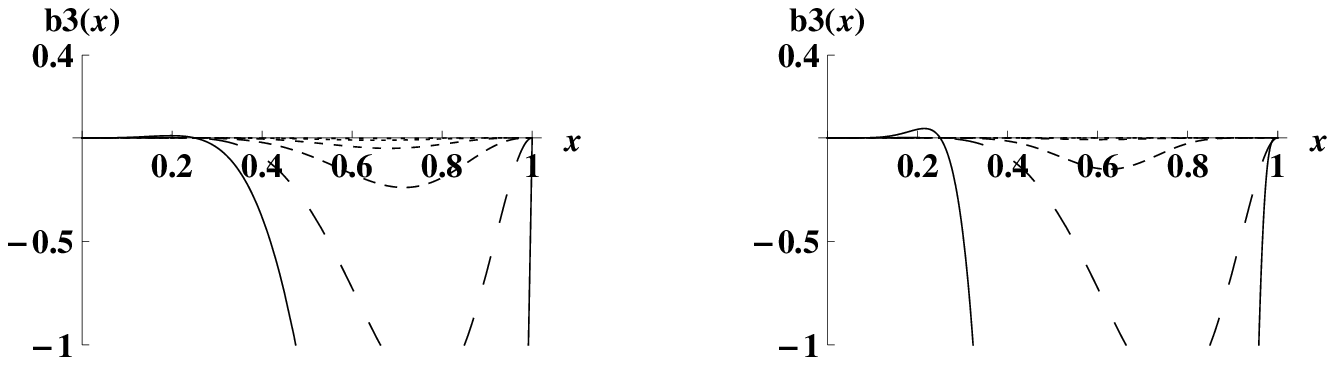, width=15cm}
\vspace{0.5cm}
\epsfig{file=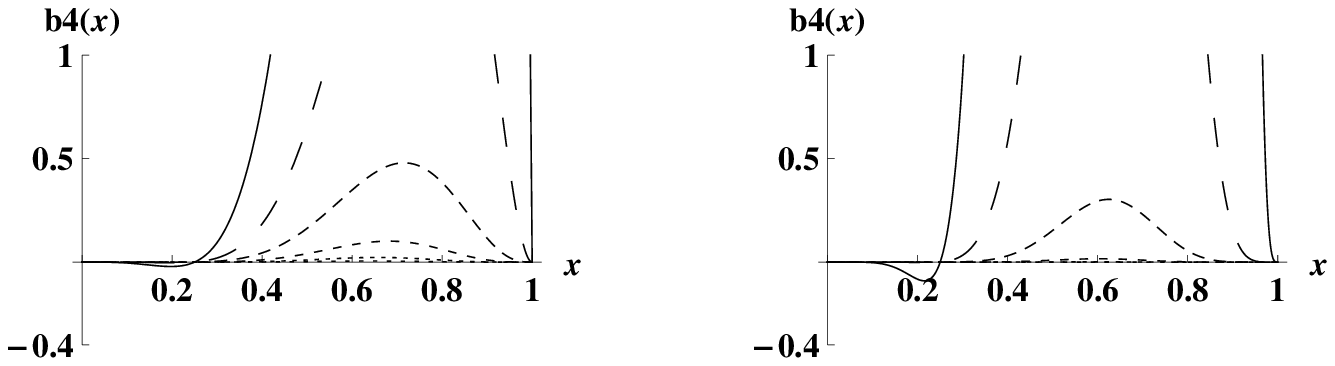, width=15cm}
\vspace{0.5cm}
\epsfig{file=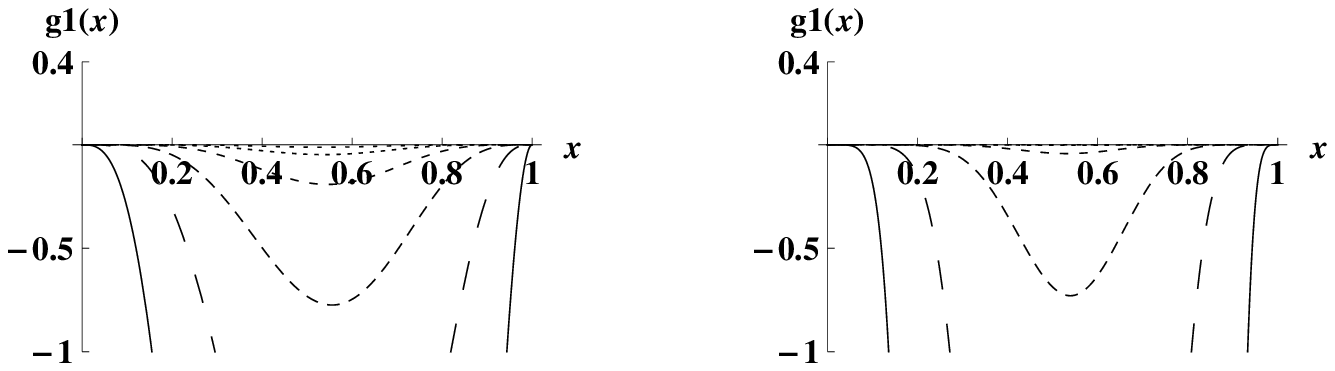, width=15cm}
\vspace{0.5cm}
\epsfig{file=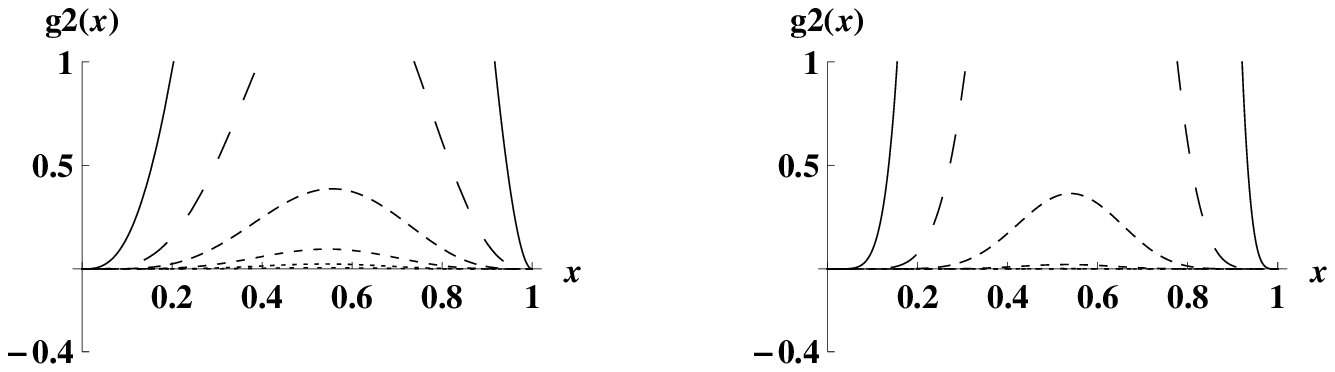, width=15cm} {\caption{\small
Structure function $b_3(x)$, $b_4(x)$, $g_1(x)$ and $g_2(x)$ for
vector mesons as a function of the Bjorken parameter scaled as
explained in the text. Left and right figures correspond to the
$\textmd{D3-D7}$-brane and the
$\textmd{D4-D8-}\overline{\textmd{D8}}$-brane models, respectively.
Different curves stand for different values of $\ell > 0$, we
display the lowest values of it. The curve sizes increase with
$\ell$. Notice that $t$ has been neglected.}}
\label{Figb4vectormesons}
\end{center}
\end{figure}

Also, we obtain the following relations: $2 b_3 = -b_4$, $2 g_2 = -
g_1$, $b_2=3 F_2$ and $b_1 =3 F_1$, which we regard as predictions
of both holographic dual models in the $t\rightarrow 0$ limit.  In
particular the relation $b_1 = 3 F_1$ is in agreement with the
expectations for a spin-one hadron since $b_1 \simeq {\cal
{O}}(F_1)$, as in the case of the $\rho$-meson, consisting of
relativistic spin-half constituents \cite{Hoodbhoy:1988am}. We would
like to point out the fact that, on the one hand, within the
supergravity approximation the structure functions do not reveal the
partonic structure of the hadron (since there is a missing
$x$-factor in the Callan-Gross relations), and on the other hand
from $b_1 \sim {\cal {O}}(F_1)$ one may infer that the meson is
composed by relativistic spin-half constituents. A possible
explanation for this behaviour is that, as suggested in
\cite{Polchinski:2002jw}, at large 't Hooft coupling all partons are
wee since parton evolution is so rapid that the probability of
finding a parton with a substantial fraction of the incoming
momentum of the initial hadronic state is vanishingly small. So,
this explains the modified Callan-Gross relations. Now, at finite
$N$ there is an argument to think of the parent hadron as surrounded
by a cloud of other hadrons, and therefore the lepton should be
scattered by one of the hadrons in the cloud. So, perhaps this is
what $b_1 =3 F_1$ indicates in the strongly coupled and large
Bjorken parameter regimes. We do not have a completely satisfactory
answer for this apparent contradiction between no partons seen from
one type of relations and the existence of relativistic components
of spin-1/2, which is inferred from the other relation. Also, we
should notice that from OPE analysis \cite{Hoodbhoy:1988am} it is
expected that $F_1$, $F_2$, $b_1$, $b_2$ and $g_1$ are twist-two
structure functions, while the rest should not receive contributions
at leading twist. Thus, this seems to be a sort of discrepancy with
respect to our results where all the structure functions are of
comparable order of magnitude. Moreover, in each model all the
structure functions have a similar dependence of momentum squared,
{\it i.e.} $q^{-2(\tau_{p}-1)}$, where the twist $\tau_{p}$ is
$\nu+1$ in our notation (note that $\nu$ is different in each
model). It would be very interesting to understand these
differences, in order to establish whether the considered
holographic dual models are good enough as to describe the internal
structure of realistic vector mesons, including both the polarized
and the unpolarized contributions.

Extensions of this work can be done to more general holographic dual
models. In particular, it should be straightforward to apply these
ideas to the $\textmd{D4-D6-}\overline{\textmd{D6}}$ model
\cite{Kruczenski:2003uq}, as well as to extend the models considered
here to the non-Abelian cases, {\it i.e.} $N_f>1$. Also it would be
interesting to apply this approach to the case of flavoured versions
of the Klebanov-Strassler solution \cite{Klebanov:2000hb}, as in the
cases of \cite{Ouyang:2003df} and \cite{Kuperstein:2004hy}. Likely,
flavoured versions of the non-supersymmetric models presented in
\cite{Kuperstein:2003yt} and \cite{Schvellinger:2004am} might be
developed and DIS studied in these cases. Similarly, these ideas
could likely be also extended to the Maldacena-N\'u\~nez solution
\cite{Maldacena:2000yy} with the addition of matter fields in the
fundamental representation \cite{Nunez:2003cf}. For all these models
it would be very interesting to unveil how the DIS structure of
spin-one mesons is manifested.

~

\centerline{\large{\bf Acknowledgments}}

~

We thank Jos\'e Goity, Mart\'{\i}n Kruczenski and Carlos N\'u\~nez
for critical reading of the manuscript, interesting comments and
discussions. The work of E.K. is supported by a doctoral fellowship
from the ANPCyT-FONCyT Grant PICT-2007-00849. This work has been
partially supported by the CONICET, the ANPCyT-FONCyT Grant
PICT-2007-00849, and the PIP-2010-0396 Grant.

\end{document}